\documentclass[aps,prb,twocolumn,groupedaddress,superscriptaddress]{revtex4-2}
\pdfoutput=1

\usepackage{amsmath}
\usepackage{comment}
\usepackage{graphicx}
\usepackage{lipsum}
\usepackage[colorlinks=true,allcolors=blue]{hyperref}
\usepackage[dvipsnames]{xcolor}
\usepackage{orcidlink}

\usepackage[caption=false]{subfig}
\newcommand{\phantomsubfloat}[1]{
    {
        \captionsetup[subfigure]{labelformat=empty}
        \subfloat[][]{#1}
    }%
}
\newcommand{\billchange}[1]{\textcolor{black}{#1}}

\makeatletter
\DeclareFontFamily{OMX}{MnSymbolE}{}
\DeclareSymbolFont{MnLargeSymbols}{OMX}{MnSymbolE}{m}{n}
\SetSymbolFont{MnLargeSymbols}{bold}{OMX}{MnSymbolE}{b}{n}
\DeclareFontShape{OMX}{MnSymbolE}{m}{n}{
    <-6>  MnSymbolE5
   <6-7>  MnSymbolE6
   <7-8>  MnSymbolE7
   <8-9>  MnSymbolE8
   <9-10> MnSymbolE9
  <10-12> MnSymbolE10
  <12->   MnSymbolE12
}{}
\DeclareFontShape{OMX}{MnSymbolE}{b}{n}{
    <-6>  MnSymbolE-Bold5
   <6-7>  MnSymbolE-Bold6
   <7-8>  MnSymbolE-Bold7
   <8-9>  MnSymbolE-Bold8
   <9-10> MnSymbolE-Bold9
  <10-12> MnSymbolE-Bold10
  <12->   MnSymbolE-Bold12
}{}

\let\llangle\@undefined
\let\rrangle\@undefined
\DeclareMathDelimiter{\llangle}{\mathopen}%
                     {MnLargeSymbols}{'164}{MnLargeSymbols}{'164}
\DeclareMathDelimiter{\rrangle}{\mathclose}%
                     {MnLargeSymbols}{'171}{MnLargeSymbols}{'171}
\makeatother

\begin{document}


\title{Shuttling Majorana zero modes in disordered and noisy topological superconductors}


\author{Bill P. Truong \orcidlink{0000-0002-5396-6808}}
	\email[]{bill.truong@mail.mcgill.ca}
	\affiliation{Department of Physics, McGill University, Montr\'{e}al, Qu\'{e}bec, Canada H3A 2T8}

\author{Kartiek Agarwal \orcidlink{0000-0002-3217-3251}}
	\affiliation{Department of Physics, McGill University, Montr\'{e}al, Qu\'{e}bec, Canada H3A 2T8}
	\affiliation{Material Science Division, Argonne National Laboratory, Lemont, Illinois 60548, USA}

\author{T. Pereg-Barnea \orcidlink{0000-0001-5465-629X}}
	\affiliation{Department of Physics, McGill University, Montr\'{e}al, Qu\'{e}bec, Canada H3A 2T8}
	\affiliation{ICFO--Institut de Ci\`{e}ncies Fot\`{o}niques, The Barcelona Institute of Science and Technology, Castelldefels, Barcelona, Spain 08860}


\date{\today}

\begin{abstract}
The braiding of Majorana zero modes (MZMs) forms the fundamental building block for topological quantum computation. Braiding protocols which involve the physical exchange of MZMs are typically envisioned on a network of topological superconducting wires. An essential component of these protocols is the transport of MZMs, which can be performed by using electric gates to locally tune sections of the wire between topologically trivial and non-trivial phases. In this work, we numerically simulate this transport by tuning a \billchange{single section} of a superconducting wire which contains either disorder (uncorrelated and correlated) or noise. We focus on the impact of these additional effects on the diabatic error, which describes unwanted transitions between the ground state and excited states. We show that the behavior of the average diabatic error is predominantly controlled by the statistics of the minimum bulk energy gap which is suppressed in the presence of disorder. The increase in diabatic error can be several orders of magnitude and is most deleterious when the disorder correlation length is a finite fraction of the \billchange{transport distance} and negligible when these lengths are far apart. In the presence of noise, the diabatic error is significantly enhanced due to optical transitions which depend on the minimum bulk energy gap as well as the frequency modes present in the noise. The results presented here serve to further characterize the diabatic error in disordered and noisy settings, which are important considerations in practical implementations of physical braiding schemes. 
\end{abstract}


\maketitle

\section{Introduction} \label{sec:intro}
Majorana zero modes (MZMs) have garnered significant interest in recent times because of their non-Abelian exchange statistics and topological protection, their potential for being experimentally realizable and their application toward fault-tolerant quantum computation \cite{Kitaev2001,Ivanov2001,Kitaev2003,Nayak2008,DasSarma2015}. Quantum information may be stored in a degenerate ground-state subspace constructed using fermions formed from MZMs. Quantum gates acting within this subspace require MZMs to be braided while the topological protection which they are endowed ensures robustness against decoherence \cite{Ivanov2001,Nayak2008}. MZMs are predicted to emerge in topological superconductors as edge modes at the boundaries separating topologically trivial and non-trivial phases \cite{Read2000,Kitaev2001,Ivanov2001}. \billchange{Given the rarity of topological superconductivity in nature, there has been much focus on engineering platforms which provide favorable conditions to host such a superconducting phase.} Proposals for platforms in quasi-1D settings have been particularly well-studied and involve topological insulators \cite{Fu2008,Fu2009,Cook2011}, nanowires built from semiconductor-superconductor heterostructures \cite{Lutchyn2010,Oreg2010,Sau2010,Sau2010a,Alicea2010,Stanescu2011}, planar Josephson junctions \cite{Hell2017,Pientka2017,Hegde2020}, atomic spin chains deposited on superconducting substrates \cite{Choy2011,Braunecker2013,Klinovaja2013,Nadj-Perge2013,Pientka2013,Rachel2025}, and chains of quantum dots connected via superconducting leads \cite{Leijnse2012,Sau2012,Fulga2013}. \billchange{The experimental fabrication of these platforms along with methods for detecting MZMs within them have seen substantial advances in recent times. For nanowire heterostructures in particular, experimental results have been encouraging with respect to detection, most notably those which rely on a quantized differential conductance \cite{Mourik2012,Das2012,Deng2012,Lee2012,Churchill2013,Finck2013,Deng2016,Nichele2017,Deng2018,Gul2018,Vaitiekenas2020,Zhang2021,Menard2020,Heedt2021,Puglia2021,Aghaee2023,Aghaee2025a}, the fractional Josephson effect \cite{Rokhinson2012,Laroche2019}, and Coulomb blockade spectroscopy \cite{Higginbotham2015,Albrecht2016,Deng2016,Albrecht2017,Shen2018,Vaitiekenas2020,VanZanten2020} (see also recent reviews in Refs. \cite{Lutchyn2018,Flensberg2021,Yazdani2023})}. However, the unambiguous confirmation of the existence of MZMs remains outstanding in large part due to the possibility of these signatures emerging from trivial bound states. Nevertheless, experimental progress continues to be promising, which has led to work on the theoretical front to focus on a range of topics related to future applications. In particular, there is a focus on the manipulation of MZMs toward their eventual braiding in practical architectures.

Various protocols for braiding MZMs have been proposed and involve the physical shuttling of MZMs through networks of superconducting wires \cite{Alicea2011,Clarke2011, Halperin2012,Bauer2019}, the direct control of couplings between MZMs \cite{Flensberg2011, Sau2011,vanHeck2012,Hyart2013,Aasen2016,Bauer2019,Malciu2018,Martin2020,min2022dynamical}, and measurement-based schemes \cite{Bonderson2008,Bonderson2013,Vijay2016,Karzig2017,Plugge2017}. As MZMs are dynamically manipulated in any braiding protocol, there always exists the possibility of nonadiabatic transitions occurring between the ground-state subspace and the excited states of a given system. These transitions are undesirable in the context of quantum computation as they lead to decoherence and are therefore regarded as a source of error. The transition probability is often referred to as the diabatic error. These errors are generally suppressed as long as protocols are performed over time-scales which greatly exceed those of the inverse bulk energy gap. However, this may be challenging to achieve in practical settings which typically demand that protocols be performed over relatively fast times. The characterization of the diabatic error is therefore essential for these situations. 

The diabatic error has been comprehensively examined in previous works for braiding protocols \cite{Cheng2011,Karzig2015,Amorim2015,Knapp2016,Rahmani2017,Sekania2017,Zhang2019,Nag2019,Harper2019,Sanno2021,Maciazek2023,Xu2023,Mascot2023,Boross2024,Peeters2024,Hodge2025} as well as for the transport of MZMs \cite{Scheurer2013,Karzig2015_2,Bauer2018,Conlon2019,Coopmans2021,Xu2022,Truong2023,Wang2024,Sahu2024,Pandey2025}, which is a core component of physical braiding schemes. \billchange{In this present work, we extend upon the study in Ref. \cite{Truong2023} by focusing on the transport of MZMs in a topological superconducting wire containing either disorder or noise, which inevitably arise in realistic platforms}. Motivated by previous works in related contexts \cite{Knapp2016,Mishmash2020,Boross2022,Boross2024}, we study the influence of these additional effects on the behavior of the diabatic error for transport and identify key parameters or quantities which substantially impact the error. The transport protocol that we consider involves tuning a \billchange{single section} of the wire in a controlled manner between the topologically trivial and nontrivial phases via an electric gate which adjusts the local chemical potentials. \billchange{This protocol can be viewed as a ``single key'' limit of the ``piano key'' transport protocol which was proposed in Ref. \cite{Alicea2011}.} As the phase boundaries shift along the wire, the MZMs which are pinned to them are shuttled along accordingly. We implement \billchange{static disorder} on the local chemical potential to model the presence of charge impurities within the wire. Both spatially uncorrelated and Gaussian-correlated disorder are considered; a correlation length $\xi$ is introduced in the latter case and treated as a parameter. \billchange{Time-dependent noise} is studied on a separate footing and is implemented by adding a noisy signal on the chemical potential to model electromagnetic interference originating from the environment, see Ref. \cite{Paladino2014} for a review. In the main text, we study $1/f$ noise with cutoff frequencies. The case of white noise is also examined and yields similar results---these results are relegated to the Appendix \ref{app:whitenoise}.

We consider a simple protocol where an MZM is transported across a superconducting wire by uniformly tuning the chemical potential over a \billchange{region} of size $R$. The behavior of the diabatic error accumulated from \billchange{this method of transport} in a clean wire has been well-studied and is found \cite{Bauer2018,Truong2023} to depend primarily on the following quantities: the protocol time $\tau$, the change in the chemical potential $\mu_{0}$, and the minimum energy gap $\Delta_{\mathrm{m}}$ between the ground state and first excited state with the same parity. For transport in a disordered wire, we find that here as well, the error is largely governed by changes in the minimum gap $\Delta_{\mathrm{m}}$. We demonstrate that minimum gap statistics can correctly predict the average behavior of the error for either uncorrelated or correlated disorder. In particular, for \billchange{$\xi/R \rightarrow 0$}, the results for the error and the minimum gap agree with that of the uncorrelated case, while the limit \billchange{$\xi/R \rightarrow 1$} behaves like the clean case, as can be expected. Remarkably, the suppression of the minimum gap is maximum for a value of \billchange{$\xi/R \sim 0.1$} and is largely independent of the \billchange{transport distance} $R$. In general, one may expect that increasing $\xi$ allows for the nucleation of subgap states, which can lead to reduced coherence. The finding that the minimum gap is suppressed the most for \billchange{$\xi/R \sim 0.1$} implies that \billchange{the minimum gap is obtained for states that are localized over a fraction of the transport distance.} \billchange{This suppressed minimum gap directly relates to an enhanced diabatic error, which may have implications for transport conducted using multiple wire sections, or multiple ``piano keys,'' as was studied in Ref. \cite{Truong2023}. For this kind of transport, we predict that piano keys should be designed to be either sufficiently short (such that $\xi/R \rightarrow 1$) so that the disorder effectively approaches the clean limit or sufficiently large (such that $\xi/R \rightarrow 0$) for the disorder to appear uncorrelated over the length of the key.}

For transport conducted on the noisy wire, we find that the diabatic error features two distinct behaviors for sufficiently long protocol times $\tau$---for the case where the high-frequency cutoff of the noise is lower than the minimum gap, the error achieves a constant value, while for high-frequency cutoff larger than the minimum gap, it increases linearly with $\tau$. For the former case, diabatic errors are governed by multi-photon absorption processes and the error is thus exponentially small for small spectral weight $A$ of the noise, while in the latter case, the error grows linearly in $A$. 


This paper is organized as follows. In Sec.~\ref{sec:preliminaries}, we review the features of the Kitaev chain, which we use to model a topological superconducting wire. We discuss the transport protocol including the chosen tuning of the chemical potential, the implemention of noise and disorder, and the computation of the error. In Sec.~\ref{sec:dis_uncorr}, we present our numerical results for the diabatic error and minimum gap statistics obtained from simulations of the transport protocol with uncorrelated disorder. In Sec.~\ref{sec:dis_corr}, we present analogous results obtained from simulations with correlated disorder. In Sec.~\ref{sec:1fnoise}, we present results for the diabatic error obtained from simulations which contain noise on the chemical potential tuning. Finally, in Sec.~\ref{sec:concl}, we conclude with a summary of our findings. Throughout this paper, we set $\hbar = 1$. 

\section{Preliminaries} \label{sec:preliminaries}
\subsection{Model and transport protocol} \label{sec:prelim:pianokey}
\begin{figure}[t!]
	\centering
	\includegraphics[width=0.45\textwidth]{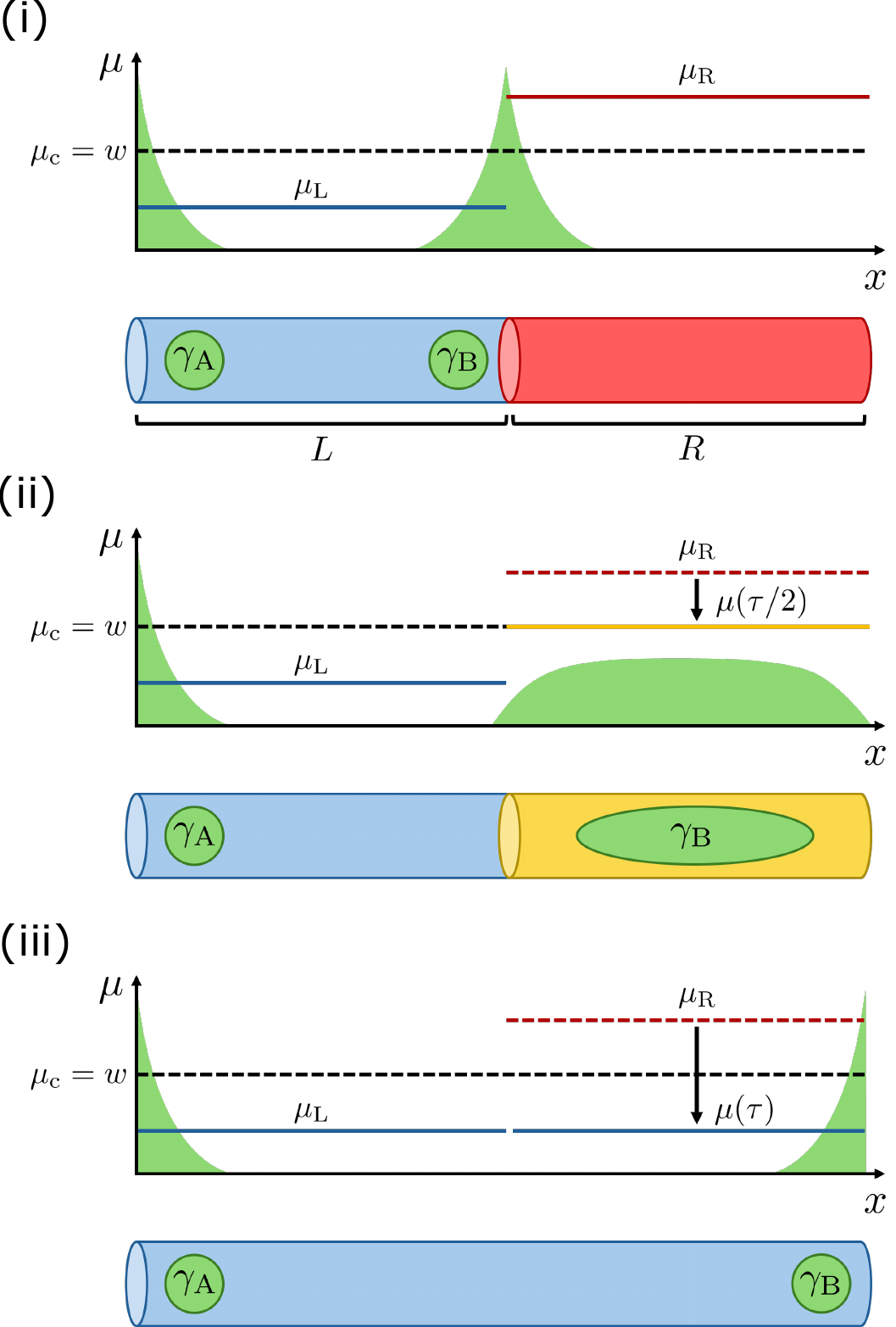}
	\phantomsubfloat{\label{fig:transportprotocol_1}}
	\phantomsubfloat{\label{fig:transportprotocol_2}}
	\phantomsubfloat{\label{fig:transportprotocol_3}}
    \caption[]{Illustration of the protocol used to transport a MZM  across a topological superconducting wire over a distance $R$ in time $\tau$ by tuning local chemical potentials. The wire contains two MZMs, $\gamma_{\mathrm{A}}$ and $\gamma_{\mathrm{B}}$, with the latter undergoing transport toward the wire's right edge. The MZM wave functions are pictorially depicted at each major step of the protocol. (i) Initial configuration of the wire, which is divided into a left section that is topologically nontrivial and a right section which is trivial. (ii) Configuration of the wire when the \billchange{right section} reaches the critical point resulting in the delocalization of $\gamma_{\mathrm{B}}$ across the \billchange{section}. (iii) End of the transport protocol with the wire being completely in the nontrivial phase and with $\gamma_{\mathrm{B}}$ located on the wire's right edge.}
  	\label{fig:transportprotocol}
\end{figure}

We use the Kitaev chain to model a topological superconducting wire \cite{Kitaev2001}. The real-space Hamiltonian for a Kitaev chain with $N$ sites in terms of electron creation and annihilation operators $c_{j}, c_{j}^{\dagger}$ is given by
\begin{align}
	H_{\text{kit}} = &- \sum_{j=1}^{N} \mu_{j}(t) c_{j}^{\dagger} c_{j} - \frac{w}{2} \sum_{j=1}^{N-1} (c_{j}^{\dagger} c_{j+1} + \text{H.c.})  
	\nonumber	
	\\
	&-\frac{\Delta_{\mathrm{SC}}}{2} \sum_{j=1}^{N-1} (c_{j} c_{j+1} + \text{H.c.}), 
	\label{eq:hamiltonian_kitaev}
\end{align}
where $\mu_{j}(t)$ is the chemical potential, $w$ is the nearest-neighbor hopping amplitude and $\Delta_{\mathrm{SC}}$ is the superconducting pairing amplitude. In anticipation of the forthcoming discussion on the transport protocol as well as the implementation of disorder and noise, the chemical potential is explicitly allowed to vary along the chain and with time.  

We now briefly review the essential features of the Kitaev chain relevant to this work. The chain features two topologically distinct phases. This can be easily demonstrated in the case where the chemical potential is constant, $\mu_{j}(t) = \mu$. The critical point separating the two phases occurs when $|\mu| = \mu_{\mathrm{c}} \equiv w$ which corresponds to a closing of the bulk energy gap. The topologically trivial phase occurs when $|\mu| > w$ and no MZMs appear in the chain. The topologically nontrivial phase occurs when $|\mu| < w$ and MZMs appear at the edges of the chain. The wave functions of the MZMs decay exponentially into the bulk. The MZMs also feature an energy splitting which depends on the overlap of their wave functions and is exponentially suppressed by the size of the chain. 

In general, different sections of the Kitaev chain can be placed in different topological phases by allowing parameters to vary from site to site. MZMs appear at the boundaries separating distinct topological phases and so moving these boundaries equates to moving the MZMs themselves. The chemical potential is the parameter chosen to control the phases of the chain sections to facilitate MZM transport. In experimental settings, the chemical potential can be tuned locally through electric gating. 

The transport protocol that we consider is identical to those studied in Refs. \cite{Bauer2018,Truong2023}. The steps of the protocol are schematically illustrated in Fig.~\ref{fig:transportprotocol}. Initially, the chain is divided into two sections. The left section contains $L$ sites and is placed in the nontrivial phase while the right section contains $R = N - L$ sites and is placed in the trivial phase. This initial configuration features two MZMs $\gamma_{\mathrm{A}}$ and $\gamma_{\mathrm{B}}$ at the left and right edges, respectively, of the left section. The MZM which is targeted for transport is $\gamma_{\mathrm{B}}$. Transport is performed by tuning the chemical potential uniformly across the right section. As the \billchange{right section} approaches the critical point, the wave function of $\gamma_{\mathrm{B}}$ delocalizes across the \billchange{section}. As the right section \billchange{is tuned} beyond the critical point and toward the end of the protocol, $\gamma_{\mathrm{B}}$ localizes on the right edge of the chain. 

The chemical potential across the chain during transport is given by
\begin{equation}
	\mu_{j}(t) = 
	\begin{cases}
		\mu_{\mathrm{L}},~\text{for}~1 \leq j \leq L,
		\\
		\mu(t),~\text{for}~ L < j \leq N,
	\end{cases}
	\label{eq:chempot_all}
\end{equation}
with the \billchange{right section's chemical potential}
\begin{equation}
	\mu(t) = \left[ 1 - f(t/\tau) \right] \mu_{\mathrm{R}} + f(t/\tau) \mu_{\mathrm{L}},
	\label{eq:chempot_tuning} 
\end{equation}
where $\mu_{\mathrm{L}} < w$ and $\mu_{\mathrm{R}} > w$ are the initial chemical potentials of the chain sections, $\tau$ is the protocol time, and $f(t/\tau)$ is the tuning function. The initial and final values for the \billchange{right section's} chemical potential are chosen symmetrically around the critical point, namely $\mu_{\mathrm{L}} = w - \mu_{0}/2$ and $\mu_{\mathrm{R}} = w + \mu_{0}/2$ where $\mu_{0}$ represents the total chemical potential change. The tuning function is such that $f(0) = 0$ and $f(1) = 1$. In our simulations, we choose a ``smooth'' function $f(t/\tau) = \sin^2(\pi t/ 2 \tau)$.

\subsection{Diabatic error} \label{sec:prelim:diaberr}
The diabatic error is the probability that a system experiences excitations between the ground-state subspace and the remaining excited states. We study the diabatic error that is accumulated by the transport protocol over a time $\tau$. Formally, the error is defined as
\begin{equation}
	\mathcal{P}(\tau) = 1 - |\langle \Omega_{\mathrm{f}} | U(\tau) | \Omega_{\mathrm{i}} \rangle|^2,
	\label{eq:diaberr}
\end{equation} 
where $| \Omega_{\mathrm{i}} \rangle$ and $| \Omega_{\mathrm{f}} \rangle$ are the initial and final many-body instantaneous ground states, respectively, while $U(\tau)$ is the time evolution operator which encodes the transport protocol. We calculate the diabatic error in Eq.~(\ref{eq:diaberr}) numerically by employing the covariance matrix method; see Appendix \ref{app:covariance} for details. The time evolution operator is calculated by discretizing time and evaluating a time-ordered product of matrix exponentials; see Appendix \ref{app:timeevol} for details. Since the Hamiltonian of the Kitaev chain conserves parity, the initial and final ground states in Eq.~(\ref{eq:diaberr}) belong to the same parity sector. When MZMs with exactly zero energy splitting are present, the ground-state subspace is two-fold degenerate with states of opposite parity $| \Omega \rangle$ and $d_{0}^{\dagger}| \Omega \rangle$ where $d_{0} = (1/2)(\gamma_{\mathrm{A}} + i \gamma_{\mathrm{B}})$. In our work, we have MZMs with nonzero but small energy splitting $\epsilon_{0}$ due to the finite chain length; in all cases that we consider, the MZM energy splitting is typically small relative to the minimum bulk energy gap--- $\epsilon_{0}/\Delta_{\mathrm{m}} \sim 10^{-3}$. Without loss of generality, we study the diabatic error from the even parity ground state. 


We now discuss the expected behavior of the diabatic error in the transport of MZMs as established in previous works \cite{Bauer2018, Truong2023}. It is understood that most of the error is attributed to transitions between the ground state and the first excited state of the same parity. These states can be cast into an effective two-level system, which can be used to capture the essential dynamics of the complete system and allows for an adequate description of the full diabatic error. The error is demonstrated to feature two main behaviors: an exponential component given by the Landau-Zener formula due to the presence of an avoided level crossing near the critical point and a power-law component originating from the continuity of the tuning function at the protocol endpoints. For the smooth tuning function used in our work, there exists an approximate analytical expression for the diabatic error given by Ref. \cite{Truong2023}:
\begin{equation}
	P \sim e^{-\tau / \tau_{\mathrm{LZ}}} + 6 \left( \frac{ \tau_{\mathrm{LZ}}}{\tau} \right)^4 \left( \frac{\Delta_{\mathrm{m}}}{2 \mu_{0}} \right)^2 \left[ 1 + \left( \dfrac{2 \mu_{0}}{\Delta_{\mathrm{m}}} \right)^2 \right]^{-4},
	\label{eq:diaberr_analytical}
\end{equation}
where $\tau_{\mathrm{LZ}}$ is the Landau-Zener time-scale:
\begin{equation}
	\tau_{\mathrm{LZ}} = \frac{\mu_{0}}{\Delta_{\mathrm{m}}^2}.
	\label{eq:tau0}
\end{equation}
For a clean chain, the minimum energy gap is predicted to be $\Delta_{\mathrm{m}} = \Delta_{R} \equiv \pi \Delta_{\mathrm{SC}}/R$. In the discussion of our results for disorder, the expression in Eq.~(\ref{eq:diaberr_analytical}) will be used in highlighting the dominant role that the minimum gap plays in controlling the diabatic error's behavior.

\subsection{Disorder implementation} \label{sec:prelim:disorder}
We disorder the chemical potential on each site of the chain as follows:
\begin{equation}
	\mu_{j}(t) \rightarrow \mu_{j}(t) + \delta \mu_{j},
	\label{eq:uncorrdisorder}
\end{equation}
where $\delta \mu_{j}$ is a time-independent disorder potential. For uncorrelated disorder, $\delta \mu_{j}$ is randomly selected from a normal distribution $\mathcal{N}(0, \sigma^2)$ with zero mean and standard deviation $\sigma$, which plays the role of the disorder strength. For correlated disorder, $\delta \mu_{j}$ is randomly selected from a multivariate normal distribution $\mathcal{N}(0, \mathbf{\Sigma})$ with zero mean and covariance matrix $\mathbf{\Sigma}$. We choose the correlations to be Gaussian and this is encoded in the matrix elements of $\mathbf{\Sigma}$ as
\begin{equation}
	\Sigma_{ij} = \llangle \delta \mu_{i} \delta \mu_{j} \rrangle = \sigma^2 \exp \left( -\frac{(i-j)^2}{2 \xi^2} \right),
	\label{eq:covariancemat_disorder}
\end{equation}
where $\xi$ is the correlation length and $\llangle \cdots \rrangle$ denotes averaging over different realizations for the disorder potential. 

\subsection{Noise implementation} \label{sec:prelim:noise}
\begin{figure*}[t!]
	\centering
	\includegraphics[width=1.0\textwidth]{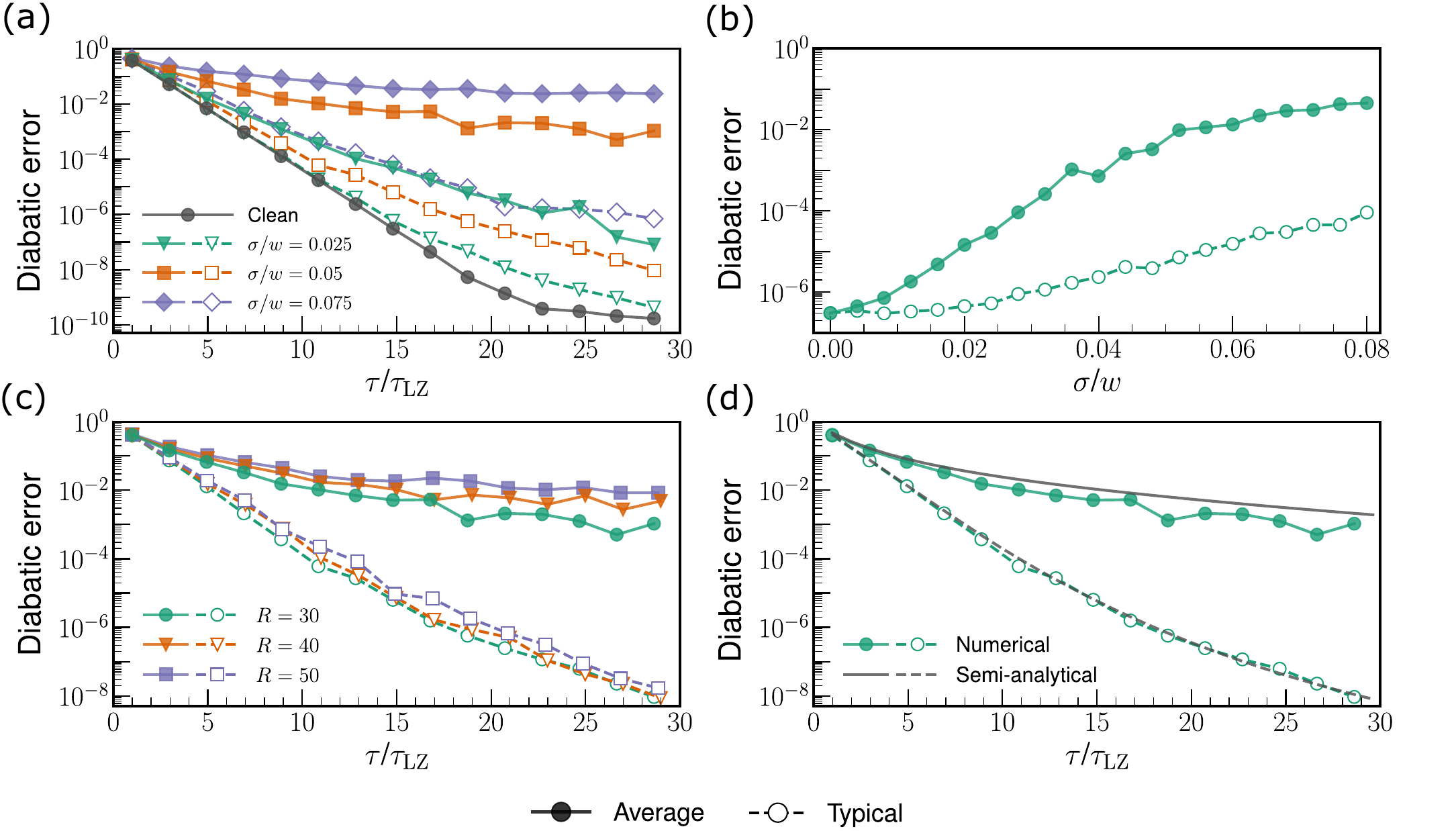}
	\phantomsubfloat{\label{fig:uncorr_diab_a}}
	\phantomsubfloat{\label{fig:uncorr_diab_b}}
	\phantomsubfloat{\label{fig:uncorr_diab_c}}
	\phantomsubfloat{\label{fig:uncorr_diab_d}}
    \caption[]{Numerical results for the diabatic error at the end of the transport protocol simulated in a chain containing uncorrelated disorder on the chemical potential. Results for the average error are displayed as solid markers with solid lines while those for the typical error are displayed as open markers with dashed lines. (a) Diabatic error versus protocol time for select disorder strengths. (b) Diabatic error versus disorder strength with protocol time $\tau/\tau_{\mathrm{LZ}} = 15$. (c) Diabatic error versus protocol time for select \billchange{transport distances} $R$ with disorder strength $\sigma/w = 0.05$. (d) Comparison between the numerical result for the diabatic error with disorder strength $\sigma/w = 0.05$ and the corresponding semianalytical expression which uses Eq.~(\ref{eq:diaberr_analytical}), actual values of the minimum gap extracted from simulations, and averaging. Unless otherwise specified, the default chain parameters are $L = R = N/2 = 30$ sites, $\mu_{0} = 0.4$ meV, $w = 3$ meV, and $\Delta_{\mathrm{SC}} = 0.6$ meV. For each result, averaging is performed over $500$ simulations with different disorder realizations.}
  	\label{fig:uncorr_diab}
\end{figure*}
We include noise on the chemical potential as follows:
\begin{equation}
	\mu(t) \rightarrow \mu(t) + \delta \mu(t),
	\label{eq:noiseimplement}
\end{equation}
where $\delta \mu(t)$ is the \billchange{time-dependent} noise signal added on top of $\mu(t)$ which tunes \billchange{the right section's} chemical potential according to the protocol. The noise signal obeys a power spectral density $S(\omega)$ which we have the freedom to specify. Following Ref. \cite{Mishmash2020}, we outline the procedure used to generate noise signals corresponding to a given $S(\omega)$. The essence of this procedure is the generation of a white noise signal which is modified in frequency space before being transformed back into temporal space. To this end, we first consider discrete time points $t = t_{n}$ such that $0 \leq t_{n} \leq T_{\mathrm{N}}$ where $T_{\mathrm{N}}$ is the maximum sample time and $dt = t_{n+1} - t_{n}$ is the time step. A white noise signal $\delta \mu_{\mathrm{WN}}(t_{n})$ is generated by selecting values from a normal distribution $\mathcal{N}(0,1)$ with zero mean and unit standard deviation. Performing a Fourier transform on this signal yields
\begin{equation}
	\delta \tilde{\mu}_{\mathrm{WN}}(\omega_{k}) = \frac{1}{\sqrt{N_{\mathrm{N}}}} \sum_{n = 1}^{N_{\mathrm{N}}} e^{-i \omega_{k} t_{n}} \delta \mu_{\mathrm{WN}}(t_{n}),
	\label{eq:ft_basenoise}
\end{equation}
where $N_{\mathrm{N}} = T_{\mathrm{N}}/dt$ is the total number of sample time points and $\omega_{k}$ are discrete frequencies. Equation (\ref{eq:ft_basenoise}) is modified using the power spectral density $S(\omega_{k})$ as follows:
\begin{equation}
	\delta \tilde{\mu}(\omega_{k}) = \sqrt{\frac{S(\omega_{k})}{dt}} \delta \tilde{\mu}_{\mathrm{WN}}(\omega_{k}).
	\label{eq:ft_noise}
\end{equation}
The noise signal of interest $\delta \mu(t)$ can be constructed by taking the inverse Fourier transform of Eq.~(\ref{eq:ft_noise}). A key quantity is the noise strength, or power, which can be calculated in the usual way by summing the power spectral density:
\begin{equation}
	\sigma^2 = \llangle \delta \mu(t_{n})^2 \rrangle = \frac{1}{T_{\mathrm{N}}} \sum_{k = 1}^{N_{\mathrm{N}}} S(\omega_{k}).
	\label{eq:ft_power}
\end{equation}
In our work, we focus on the specific case of $1/f$ noise which is widely believed to reflect the behavior of underlying two-level fluctuators in physical settings \cite{Paladino2014}. The power spectral density is
\begin{equation}
	S(\omega_{k}) = 
	\begin{cases}
		\dfrac{A}{\omega_{k}}, &\text{for}~ \omega_{\mathrm{l}} \leq \omega_{k} \leq \omega_{\mathrm{h}},
		\\
		0, &\text{otherwise},
	\end{cases}
	\label{eq:noise1f}
\end{equation}
where $A$ is the noise amplitude, $\omega_{\mathrm{l}}$ is the low frequency cutoff, and $\omega_{\mathrm{h}}$ is the high-frequency cutoff. Though the main text is concerned with $1/f$ noise, we remark that we have also considered the effects of using white noise; see Appendix \ref{app:whitenoise}.

\section{Uncorrelated disorder} \label{sec:dis_uncorr}
\begin{figure*}[t!]
	\centering
	\includegraphics[width=1.0\textwidth]{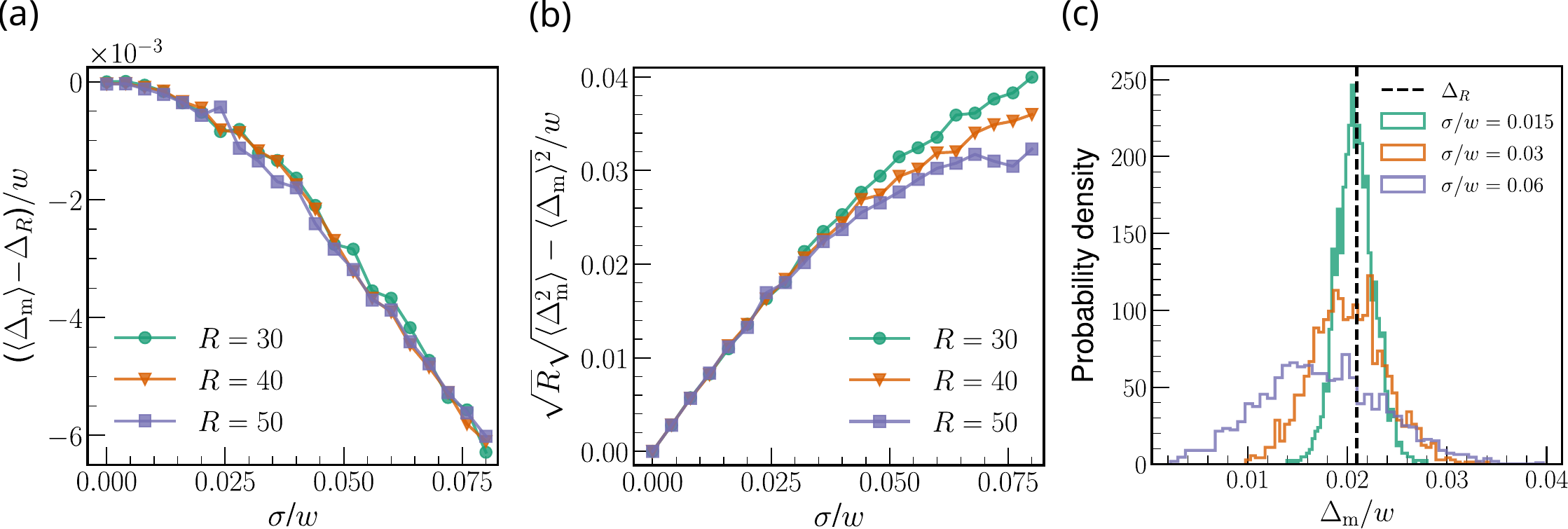}
    \phantomsubfloat{\label{fig:uncorr_mingap_a}}
	\phantomsubfloat{\label{fig:uncorr_mingap_b}}
	\phantomsubfloat{\label{fig:uncorr_mingap_c}}
	\caption[]{Numerical results for the minimum gap statistics of the transport protocol simulated in a chain with uncorrelated disorder on the chemical potential. (a) Average minimum gap versus disorder strength for select \billchange{transport distances} $R$. The clean minimum gap $\Delta_{R}$ is subtracted from each result to demonstrate the absence of any $R$-dependent contributions from the disorder. (b) Standard deviation of the minimum gap versus disorder strength for select $R$. The results are scaled with respect to $\sqrt{R}$ to demonstrate the $\sim 1/\sqrt{R}$ dependence for weak disorder. (c) Normalized probability densities corresponding to select disorder strengths. The clean minimum gap value $\Delta_{R}$ is highlighted by a vertical dashed black line. Unless otherwise specified, the default chain parameters are identical to those cited in Fig.~\ref{fig:uncorr_diab} and in the main text. For panels (a) and (b), each result is obtained by averaging over $1500$ simulations with different disorder realizations. For panel (c), each density is constructed by sampling $1500$ values of the minimum gap.}
  	\label{fig:uncorr_mingap}
\end{figure*}

\begin{figure*}[th]
	\centering
	\includegraphics[width=1.0\textwidth]{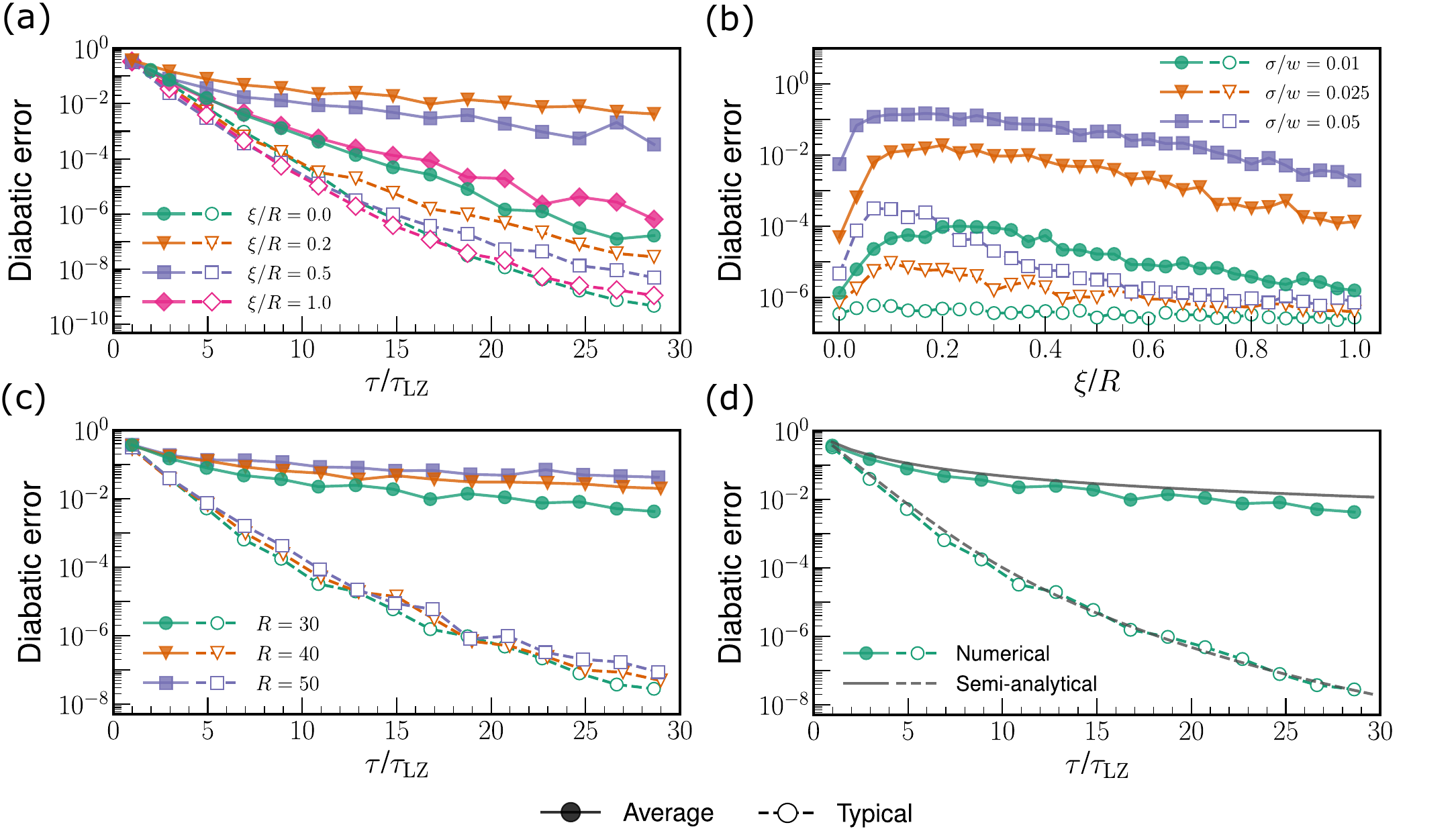}
	\phantomsubfloat{\label{fig:corr_err_a}}
	\phantomsubfloat{\label{fig:corr_err_b}}
	\phantomsubfloat{\label{fig:corr_err_c}}
	\phantomsubfloat{\label{fig:corr_err_d}}
    \caption[]{Numerical results for the diabatic error at the end of the transport protocol simulated in a chain containing Gaussian-correlated disorder on the chemical potential. Results for the average error are displayed as solid markers with solid lines while those for the typical error are displayed as open markers with dashed lines. (a) Diabatic error versus protocol time for select correlation lengths with disorder strength $\sigma/w = 0.025$. (b) Diabatic error versus correlation length for select disorder strengths with protocol time $\tau/\tau_{\mathrm{LZ}} = 15$. (c) Diabatic error versus protocol time for select \billchange{transport distances} $R$ with disorder strength $\sigma/w = 0.025$ and correlation length $\xi/R = 0.2$. (d) Numerical result for the diabatic error with disorder strength $\sigma/w = 0.025$ and correlation length $\xi/R = 0.2$ in comparison to the corresponding semianalytical expression. Unless otherwise specified, the default chain parameters are identical to those cited in Fig.~\ref{fig:uncorr_diab} and in the main text. Averaging is performed over $500$ simulations for each result.}
  	\label{fig:corr_diab}
\end{figure*}

The transport protocol is simulated on a Kitaev chain with uncorrelated disorder on the chemical potential. The diabatic error is numerically calculated using Eq.~(\ref{eq:diaberr}) and averaging is performed over 500 simulations which differ only in their disorder potential configuration. We remark that a different disorder potential is randomly generated for each simulation that is performed. We calculate the average error $\llangle \mathcal{P} \rrangle$ and the typical error through the geometric average $\exp[\llangle\log(\mathcal{P})\rrangle]$. Since the error can vary over many orders of magnitude in our work, the typical error can serve as a more useful diagnostic in this setting due to its ability to suppress outliers. Unless otherwise specified, the default chain parameters used in all simulations are $L = R = N/2 = 30$ sites, $\mu_{0} = 0.4$ meV, $w = 3$ meV, and $\Delta_{\mathrm{SC}} = 0.6$ meV. \billchange{The corresponding Landau-Zener time-scale in Eq. (\ref{eq:tau0}) is $\tau_{\mathrm{LZ}} \approx 0.067$~ns in physical units.} These parameters are identical to those used in Ref. \cite{Truong2023}. 

The main results for the diabatic error in the case of uncorrelated disorder are shown in Fig.~\ref{fig:uncorr_diab}. The filled and open symbols indicate the average and typical errors, respectively, in all plots. As illustrated in Fig.~\ref{fig:uncorr_diab_a}, the error increases with disorder strength $\sigma$ for each of the protocol times displayed. This increase can be seen more directly in Fig.~\ref{fig:uncorr_diab_b} where the error is varied against the disorder strength for a fixed protocol time. We also consider the effects of \billchange{changing the transport distance by changing the size of the right section.} These results are shown in Fig.~\ref{fig:uncorr_diab_c} for a fixed disorder strength. When the protocol time is rescaled with the Landau-Zener time-scale $\tau_{\mathrm{LZ}}$ defined in Eq.~(\ref{eq:tau0}), the errors corresponding to different transport distances $R$ are observed to strongly overlap with one another. We remark that $\tau_{\mathrm{LZ}}$ here uses the predicted value of the clean minimum gap $\Delta_{R}$ which explicitly depends on $R$. This scaling behavior of the error is present for a clean transport protocol and we demonstrate here that it survives even in a disordered setting. 

As previously discussed, the diabatic error for a clean transport protocol is primarily controlled by the protocol time $\tau$, the total change in the chemical potential $\mu_{0}$, and the minimum energy gap $\Delta_{\mathrm{m}}$. This can be seen directly from the analytical expression in Eq.~(\ref{eq:diaberr_analytical}). For a disordered transport protocol, we find that the error is most sensitive to changes in the minimum gap $\Delta_{\mathrm{m}}$ and demonstrate that it can be used to accurately predict the error's behavior through a semianalytical approach. This semianalytical approach relies on using the expression in Eq.~(\ref{eq:diaberr_analytical}) along with minimum gaps obtained numerically from simulations. The minimum gaps are calculated using $\Delta_{\mathrm{m}} = \mathrm{min}\{\epsilon_{1}(t) - \epsilon_{0}(t)\}$ where $\epsilon_{0}$ is the MZM energy and $\epsilon_{1}$ is the energy of the first excited single-particle state. The minimum value is taken over the duration of the protocol, $t \in [0, \tau]$. These actual minimum gaps are inserted into Eq.~(\ref{eq:diaberr_analytical}) for the diabatic error and both the average and geometric average of this run-by-run error is computed. We note that when only the protocol time is varied, this averaging is repeated with the same collection of actual minimum gaps. The results of this semianalytical approach are presented in Fig.~\ref{fig:uncorr_diab_d}. We observe that it yields excellent agreement with the corresponding numerical results for the average and typical diabatic error. 

Given the minimum gap's central role in computing the diabatic error, we study its statistics while varying the disorder strength. The results for these statistics are illustrated in Fig.~\ref{fig:uncorr_mingap} which shows the average, standard deviation, and probability density of the minimum gap. As illustrated in Figs.~\ref{fig:uncorr_mingap_a} and \ref{fig:uncorr_mingap_b}, the average minimum gap decreases with disorder strength while the standard deviation increases. For weak disorder, the average and standard deviation may be described using a semianalytical perturbation theory approach, see Appendix \ref{app:ptheory}. We remark that the behaviors observed here are consistent with previous works which have studied the minimum gap in static disordered settings \cite{Brouwer2011, Cai2013, Hegde2016}. These behaviors are simultaneously captured in the probability densities shown in Fig.~\ref{fig:uncorr_mingap_c}. Furthermore, we perform a scaling analysis on the average and standard deviation with respect to the transport distance $R$. For the average minimum gap, results corresponding to different $R$ are observed to strongly overlap when the clean minimum gap $\Delta_{R}$ is subtracted off of each result. Interestingly, this suggests that the disorder's contribution to the minimum gap is independent of $R$. For the standard deviation, a strong overlap of results is observed for weak disorder when the results are rescaled with respect to $\sqrt{R}$.

\section{Correlated disorder} \label{sec:dis_corr}
\begin{figure*}[th]
	\centering
	\includegraphics[width=0.95\textwidth]{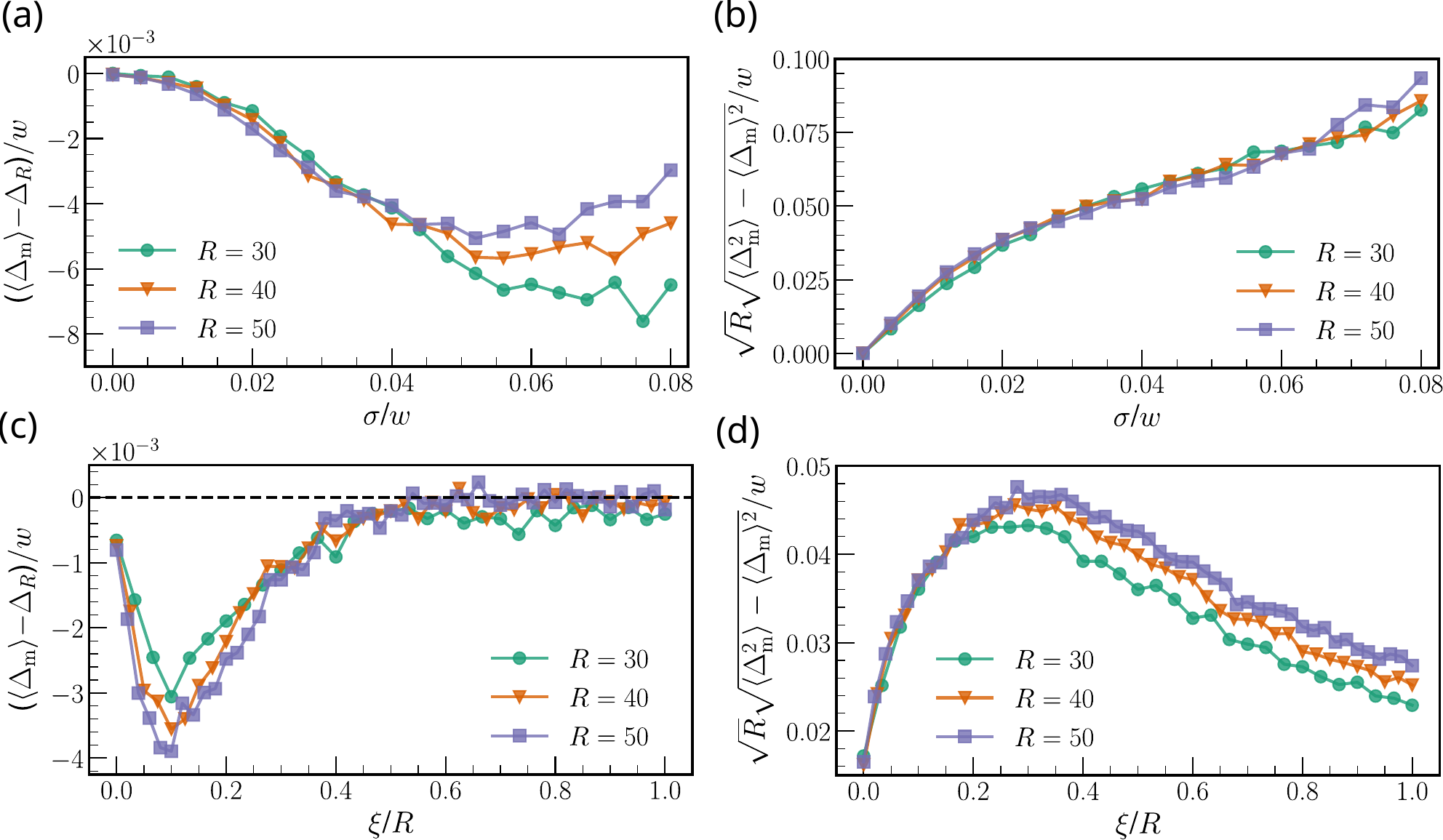}
	\phantomsubfloat{\label{fig:corr_mingap_a}}
	\phantomsubfloat{\label{fig:corr_mingap_b}}
	\phantomsubfloat{\label{fig:corr_mingap_c}}
	\phantomsubfloat{\label{fig:corr_mingap_d}}
    \caption[]{Numerical results for the minimum gap statistics of the transport protocol simulated in a chain with Gaussian-correlated disorder on the chemical potential. Results corresponding to select transport distances $R$ are displayed in each plot. (a) Average and (b) standard deviation of minimum gap versus disorder strength with correlation length $\xi/R = 0.2$. (c) Average and (d) standard deviation of minimum gap versus correlation length with disorder strength $\sigma/w = 0.025$. The horizontal dashed black line in panel (c) indicates the clean minimum gap value $\Delta_{R}$. To highlight certain scaling behaviors, $\Delta_{R}$ is subtracted off of the averages while the standard deviations are multiplied by $\sqrt{R}$. Unless otherwise specified, the default chain parameters are identical to those cited in Fig.~\ref{fig:uncorr_diab} and in the main text. Averaging is performed over $1500$ simulations for each result.}
  	\label{fig:corr_mingap}
\end{figure*}

\begin{figure*}[th]
	\centering
	\includegraphics[width=1.0\textwidth]{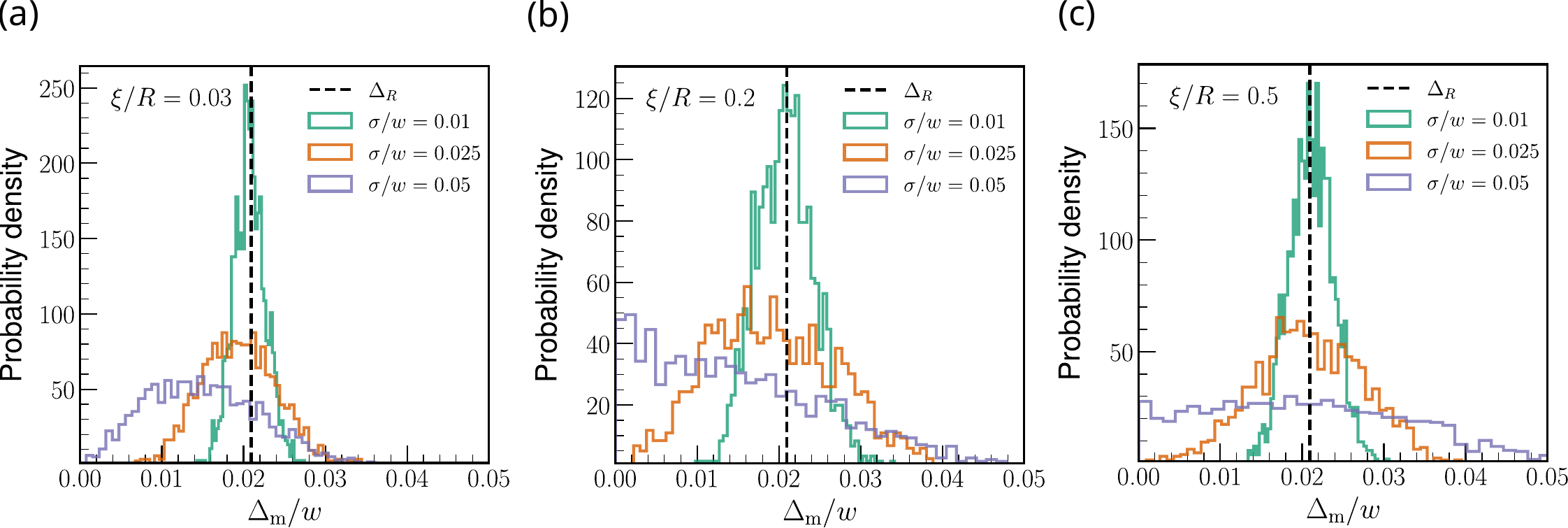}
	\phantomsubfloat{\label{fig:corr_mingapdistr_a}}
	\phantomsubfloat{\label{fig:corr_mingapdistr_b}}
	\phantomsubfloat{\label{fig:corr_mingapdistr_c}}
    \caption[]{Normalized probability densities of the minimum gap obtained from simulations. Each plot corresponds to a particular correlation length and features densities for select disorder strengths. The clean minimum gap value $\Delta_{R}$ is highlighted in each plot by a vertical dashed black line. The default chain parameters are identical to those cited in Fig.~\ref{fig:uncorr_diab} and in the main text. Each density is constructed by numerically sampling $1500$ values of the minimum gap.}
  	\label{fig:corr_mingapdistr}
\end{figure*}

We now present our results corresponding to simulations of the transport protocol on a Kitaev chain with Gaussian-correlated disorder. The default chain parameters used in these simulations are identical to those of Sec.~\ref{sec:dis_uncorr}. As before, all averages for the diabatic error are performed over 500 simulations which differ in their disorder potential configuration.                                                     

The main results for the average diabatic error are shown in Fig.~\ref{fig:corr_diab} and focus on the dependence with the correlation length $\xi$. We remark that the error's dependence with the disorder strength is similar to the dependence found for uncorrelated disorder. In Fig.~\ref{fig:corr_err_a}, the error is observed to vary nonmonotonically with the correlation length for each of the protocol times displayed. Specifically, the error initially rises with the correlation length before eventually decreasing. This is demonstrated more clearly in Fig.~\ref{fig:corr_err_b} which shows the error as a function of the correlation length directly for a fixed protocol time. The effect of disorder with short correlations on the average error is highlighted here as it can be substantial, leading to an enhancement of the error by several orders of magnitude. Figure~\ref{fig:corr_err_c} examines the dependence of the error on the transport distance $R$ at a fixed disorder strength and correlation length. We again see that these results display strong overlap with each other when the protocol times are scaled against the Landau-Zener time-scale $\tau_{\mathrm{LZ}}$. Finally, using the semianalytical approach to numerically compute the average and typical minimum gaps and diabatic errors, we find excellent agreement with the corresponding numerical results; see Fig.~\ref{fig:corr_err_d}.

The above results show that the size of the \billchange{right wire section} can be optimized with the knowledge of the disorder correlation length. \billchange{This has implications for transport using multiple wire sections, or multiple ``piano keys.''} We examine this aspect in more detail by studying minimum gap statistics next, varying the disorder strength and correlation length. Our results for these statistics are illustrated in Fig.~\ref{fig:corr_mingap}, which shows the average and standard deviation for various \billchange{transport distances}, and in Fig.~\ref{fig:corr_mingapdistr}, which shows the probability density. In Fig.~\ref{fig:corr_mingap_a}, the average minimum gap is varied against the disorder strength and exhibits nonmonotonic behavior. We observe that the average initially decreases with disorder strength, similar to what is seen for uncorrelated disorder. As the disorder strength is increased further, the average reverses its behavior and begins increasing. For weak disorder $\sigma/w \lesssim 0.04$, results corresponding to different \billchange{transport distances} $R$ show signs of overlap when the clean minimum gap $\Delta_{R}$ is subtracted off of each result, suggesting that the disorder contribution to the minimum gap is independent of $R$ only in this regime. In Fig.~\ref{fig:corr_mingap_b}, the standard deviation is shown to increase with the disorder strength. When scaled against $\sqrt{R}$, results corresponding to different $R$ strongly overlap, suggesting that the standard deviation continues to scale as $\sim 1/\sqrt{R}$.

The results for the minimum gap statistics as a function of the correlation length are illustrated in Figs.~\ref{fig:corr_mingap_c} and \ref{fig:corr_mingap_d}. The average minimum gap in Fig.~\ref{fig:corr_mingap_c} features nonmonotonic behavior since it initially decreases with the correlation length until some minimum value at $\xi/R \sim 0.1$ before increasing towards the clean minimum gap $\Delta_{R}$ for longer correlations. For short correlations, the disorder is more likely to induce isolated potential wells and barriers into the system which tend to localize single-particle, low-energy excited states. These localized states can contribute to a further suppression of the average minimum gap leading to increased diabatic errors. However, for long correlations on the order of the \billchange{transport distance}, the disorder effectively amounts to a constant shift in the chemical potential throughout the \billchange{right wire section}, and the error is close to that obtained in the clean case. Importantly, we see that the minimum gap is most suppressed at finite $\xi/R \sim 0.1$, independently of the \billchange{transport distance}. \billchange{We have verified that this suppression still occurs at $\xi/R \sim 0.1$ when model parameters such as the hopping amplitude and disorder strength are varied. For different values of $\Delta_{\mathrm{SC}}$, we observe the maximum suppression occuring for a range of values $\xi/R \sim 0.1$-$0.2$}. Nevertheless, \billchange{the scaling of $\xi$ with respect to $R$ for maximum suppression} suggests that the low energy excited states that play a key role in determining the diabatic error must be sufficiently delocalized over the \billchange{right wire section}. \billchange{For transport using multiple wire sections, we predict that each section} should thus be optimized to avoid being of similar length scales to the disorder correlation length since the effect on the diabatic error can be extremely large; see Fig.~\ref{fig:corr_err_b}.


For the standard deviation in Fig.~\ref{fig:corr_mingap_d}, we observe similar nonmonotonic behavior. The standard deviation initially rises toward a maximum value before eventually falling with the correlation length. For long correlations, the standard deviation is expected to tend toward zero as this regime coincides with the system's return to the clean case. We also check the scaling with respect to $\sqrt{R}$ and show that the results feature some overlap for short correlations $\xi/R \lesssim 0.2$.

Probability densities for the minimum gap are displayed in Fig.~\ref{fig:corr_mingapdistr}. For short correlations, as shown in Fig.~\ref{fig:corr_mingapdistr_a}, the densities resemble those of the uncorrelated disorder case. As the correlations become longer, we see in Fig.~\ref{fig:corr_mingapdistr_b} that the densities shift to smaller minimum gap values and broaden. For strong disorder, the density is seen to change quite drastically and peaks closer to a minimum gap value of zero. The accumulation of the density close to zero gap implies a proliferation of localized excited states with lower energies. Referring to the corresponding diabatic error results with the same disorder strength in Fig.~\ref{fig:corr_err_c}, though these localized states can lead to a dramatic increase of the error, transport may not be immediately destroyed, indicating that the protocol is still capable of operating in this regime. Nevertheless, comparing Figs.~\ref{fig:corr_mingapdistr_a}, \ref{fig:corr_mingapdistr_b}, \ref{fig:corr_mingapdistr_c}, it is clear that the minimum gap is suppressed most strongly for $\xi/R \sim 0.1$ independently of the \billchange{transport distance} $R$ and suggests that the relevant low energy states that control the error are delocalized over the \billchange{changing wire section.}


\section{$1/f$ noise} \label{sec:1fnoise}
\begin{figure*}[t]
	\centering
	\phantomsubfloat{\label{fig:noise_diab_a}}
	\phantomsubfloat{\label{fig:noise_diab_b}}
	\phantomsubfloat{\label{fig:noise_diab_c}}
	\phantomsubfloat{\label{fig:noise_diab_d}}
	\includegraphics[width=1.0\textwidth]{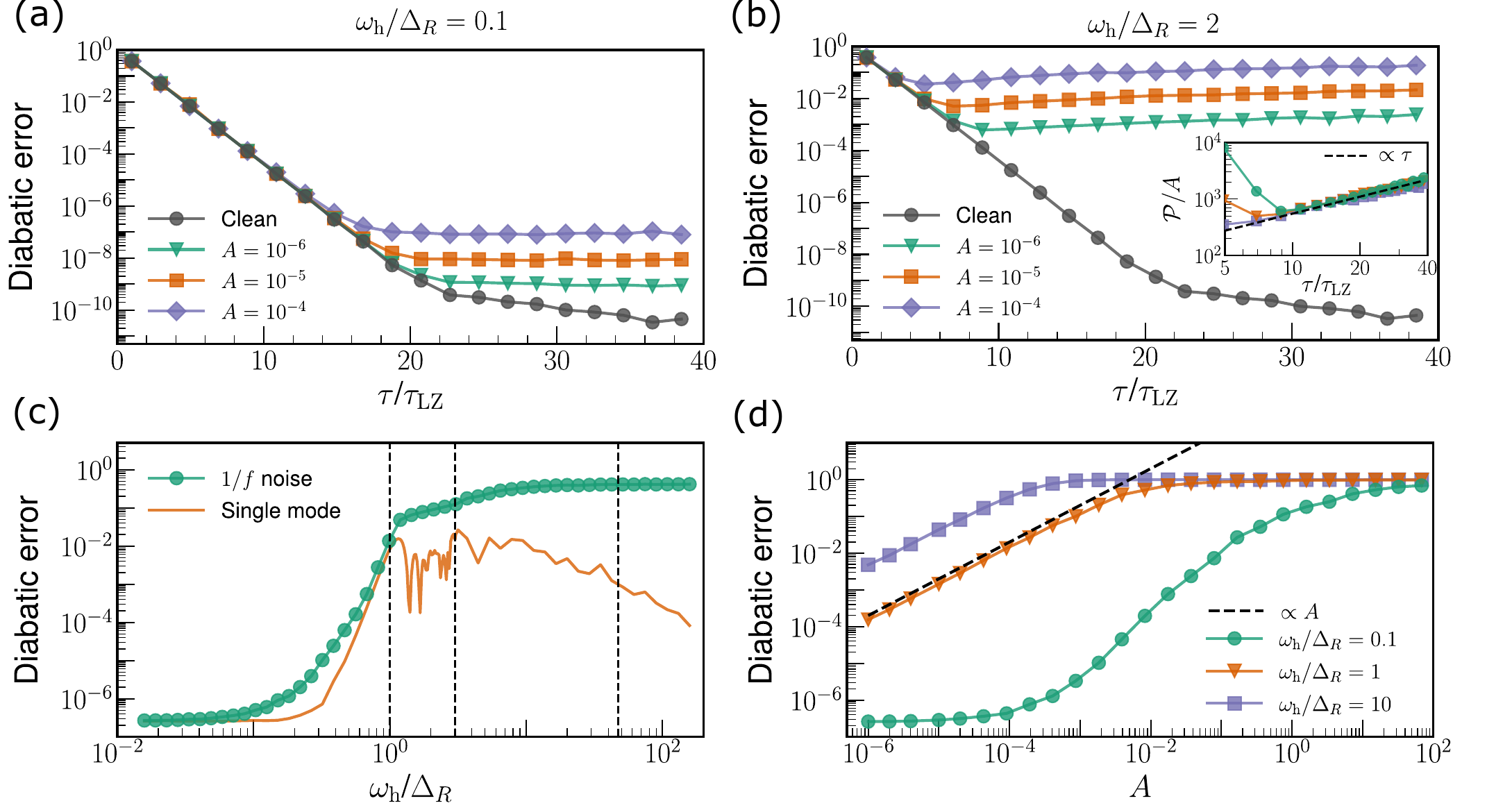}
    \caption[]{Numerical results for the average diabatic error at the end of the transport protocol simulated with $1/f$ noise added to the chemical potential tuning. (a) Diabatic error versus protocol time for select noise amplitudes \billchange{[in units of (meV)$^2$]} with high-frequency cutoff $\omega_{\mathrm{h}}/\Delta_{R} = 0.1$ which is below the minimum gap. (b) Same as panel (a) but with high-frequency cutoff $\omega_{\mathrm{h}}/\Delta_{R} = 2$ which is above the minimum gap. The inset illustrates the linear rise in the average diabatic error for sufficiently long protocol times. (c) Diabatic error versus high-frequency cutoff in the case of $1/f$ noise and frequency in the case of a single-mode with protocol time fixed to $\tau/\tau_{\mathrm{LZ}} = 15$ for both. The $1/f$ noise amplitude is \billchange{$A = 10^{-4}$ (meV)$^2$} while the single mode amplitude is \billchange{$A_{\mathrm{SM}} = 5 \times 10^{-3}$ meV}. Here, in order from left to right, the dashed black lines correspond to the clean minimum gap, the clean maximum gap, and the hopping amplitude. (d) Diabatic error versus noise amplitude for select high-frequency cutoffs with protocol time $\tau/\tau_{\mathrm{LZ}} = 15$. Unless otherwise specified, the default chain parameters are identical to those cited in Fig.~\ref{fig:uncorr_diab} and in the main text. Averaging is performed over $500$ simulations for each result. For the single-mode results, averaging is performed over the phase.}
  	\label{fig:noise_diab}
\end{figure*}

We now present results corresponding to simulations of the transport protocol with $1/f$ noise added to the tuning of the \billchange{right section's} chemical potential. The default parameters used in these simulations are identical to those used in Sec.~\ref{sec:dis_uncorr}. All averages are performed over $500$ simulations which differ only in their noise signals. 

The main results for the diabatic error are illustrated in Fig.~\ref{fig:noise_diab}. We remark that only the average error is shown in each plot. Here, we find that the average and typical errors nearly coincide, indicating a tight spread in the results. Figures \ref{fig:noise_diab_a} and \ref{fig:noise_diab_b} show the error versus the protocol time for high-frequency cutoffs $\omega_{\mathrm{h}} < \Delta_{R}$ and $\omega_{\mathrm{h}} > \Delta_{R}$, respectively. Two different behaviors emerge for long protocol times $\tau \gg \tau_{\mathrm{LZ}}$. \billchange{For $\omega_{\mathrm{h}} > \Delta_{R}$, the error increases linearly with the protocol time $\tau$, while for $\omega_{\mathrm{h}} < \Delta_{R}$, it becomes constant in this long-time limit. We can understand the linear increase in $\tau$ for $\omega_{\mathrm{h}} > \Delta_{R}$ from Fermi's golden rule---the frequencies in the noise spectrum above the instantaneous gap continue to excite the system at a rate proportional to the spectral weight. We also remark that this linear behavior has been observed in a two-level Landau-Zener model with longitudinal noise \cite{Malla2017,Krzywda2020}, which can be used as a simple model for the dynamics studied here. For $\omega_{\mathrm{h}} < \Delta_{R}$, there is no spectral weight above the gap at any instant of time, and the excitations occur due to the intrinsic properties of the Landau-Zener transition. It is known that $m$th-order discontinuities in the derivative of the driving field of the Landau-Zener transition lead to an error which goes as $\sim \tau^{-2m}$ \cite{Garrido1962,Truong2023}. Thus, it is plausible that for a discontinuous schedule (due to noise, corresponding to $m = 0$), one obtains a constant, nonzero error in the limit that $\tau \rightarrow \infty$.}

For short protocol times, the noise has little effect as the results follow the clean diabatic error closely. As well, in the case where $\omega_{\mathrm{h}} > \Delta_{R}$, the error experiences a dramatic increase for a fixed protocol time and noise amplitude. This is highlighted directly in Fig.~\ref{fig:noise_diab_c} where the error is plotted against the high-frequency cutoff. As the high-frequency cutoff $\omega_{\mathrm{h}}$ approaches the clean minimum gap $\Delta_{R}$, the noise gains additional frequency modes which become capable of inducing optical transitions into the system. This increased ability to induce optical transitions leads to the rapid increase in the error. In Fig.~\ref{fig:noise_diab_d}, the distinction between the two high-frequency cutoff regimes can be seen from the perspective of the noise amplitude. As expected, the error always increases with the noise amplitude $A$ which simply represents the noise strength, see Eq.~(\ref{eq:noise1f}). However, this increase is observed to be predominantly linear for $\omega_{\mathrm{h}} > \Delta_{R}$ and ``activated'' for $\omega_{\mathrm{h}} < \Delta_{R}$. Since the error increases linearly with $A$ in the regime $\omega_{\mathrm{h}} > \Delta_{R}$, this regime can be regarded as following Fermi's golden rule. \billchange{Moreover, this linear dependence is also captured in the behavior of the diabatic error emerging from an effective Landau-Zener model with longitudinal noise \cite{Krzywda2020}.}

It is clear that the frequency modes in the noise play a major role in the diabatic error's behavior. To demonstrate this point more directly, we consider the simple case of a single frequency mode added to an otherwise constant chemical potential tuning. This single mode takes the form
\begin{equation}
	\delta \mu_{\mathrm{SM}}(t) = A_{\mathrm{SM}} \sin(\omega t + \phi),
	\label{eq:singlemode}
\end{equation}
where $A_{\mathrm{SM}}$ is the amplitude, $\omega$ is the frequency and $\phi$ is the phase. The transport protocol is simulated with the noise signal $\delta \mu(t)$ replaced with the single mode $\delta \mu_{\mathrm{SM}}(t)$. Our results for the diabatic error in this case are illustrated in Fig.~\ref{fig:noise_diab_c} and are compared to the results for $1/f$ noise. The single mode results feature similar behavior compared to those for $1/f$ noise. The sharp rise in the diabatic error reflects the fact that the single-mode as well as any mode in the $1/f$ noise spectrum can induce optical transitions if they are within range of the clean minimum gap. 
The single-mode error decreases exponentially with frequency beyond the maximum single particle gap; these results are in agreement with the literature on heating in quantum systems with high-frequency drives~\cite{mori2016rigorous,agarwal2020dynamical,martin2022effect,Martin2020}. Concomitantly, for $1/f$ noise, we see that increasing the high-frequency cutoff beyond the maximum gap does not significantly change the diabatic error.  

\section{Discussion and conclusion} \label{sec:concl}
In this work, we study the diabatic error arising from the transport of an MZM across a superconducting wire containing disorder or noise. The diabatic error is calculated through numerical simulation of the transport protocol, which relies on dynamically tuning the chemical potential in sections of the wire through electric gating. We find that disorder affects the diabatic error principally through the minimum energy gap between the ground state and first excited state of the same parity. We demonstrate that precise knowledge of the minimum gap statistics combined with analytical expressions for the error in a clean setting can be effective in accurately predicting the behavior of the error in a disordered setting. For both uncorrelated and correlated disorder, the error increases with the disorder strength which is explained by the suppression of the average minimum gap. In the case of correlated disorder, the error features nonmonotonic behavior with respect to the disorder correlation length which closely aligns with the nonmonotonic behavior observed in the minimum gap statistics. In particular, we find that the diabatic error can vary over orders of magnitude and is largest for disorder correlation $\xi/R \sim 0.1$, where $R$ is \billchange{the transport distance}. 

In our investigation of the transport protocol with $1/f$ noise, we show that the diabatic error features two distinct long-time behaviors that chiefly depend on whether the high-frequency cutoff is below or above the clean minimum gap. In the latter case, the error is also dramatically enhanced. These effects are directly related to the ability of the noise to induce optical transitions, with the frequency cutoffs serving to control this ability through the noise's frequency modes. 

\billchange{While the focus of this work is on the diabatic error, it is important to comment on other errors that emerge with the dynamical manipulation of MZMs, particularly those which flip the fermionic parity. In realistic platforms, parity-flip errors can be caused by quasiparticle poisoning and the time-scales over which they occur has been estimated to be on the order of nanoseconds (e.g. for typical nanowire heterostructure setups \cite{Rainis2012}) to minutes (e.g. for floating wire setups in the Coulomb blockade regime \cite{Knapp2018,Karzig2021}). For the results presented in this work, transport simulations were performed with typical protocol times on the order of $\tau \sim \tau_{\mathrm{LZ}} \sim 0.1$~ns, which is generally shorter than the quoted quasiparticle time-scales. This highlights the importance of diabatic errors as they can occur before parity-flip errors. Moreover, the likelihood of parity-flip errors can be distilled by examining a clean system under sufficiently low temperature. Here, the only unpaired, localized states come from the MZMs. The tunneling of a quasiparticle into these localized states occurs with significantly less probability compared to the bulk states. A further examination on the subject of quasiparticle poisoning is beyond the scope of this work.}

\billchange{We briefly discuss the experimental feasibility of the transport protocol. The protocol relies on the local tuning of the chemical potential along a nanowire which can be achieved via nearby electric gates. Experiments on InAs-Al and InSb-Al/Pt nanowire heterostructures with lengths $\sim0.5$-$2~\mu$m have demonstrated that these gates can be fabricated with lengths as small as $\sim50$-$100$~nm using electron beam lithography and evaporation \cite{DeMoor2018,Vaitiekenas2018,Mazur2022,Wang2022,Mazur2024}. In particular, recent experiments on quantum dot-superconductor platforms used to build few-site Kitaev chains have demonstrated that these gates can be placed side-by-side in a keyboard configuration \cite{Dvir2023,Bordin2024,Zatelli2024,Bordin2025}. The advances in gate fabrication seen thus far in these experiments are therefore encouraging with respect to a potential future realization of the transport protocol.}

\billchange{The applicability of this work is mainly toward braiding that is performed on networks of nanowires which contain T-junctions. Here, the transport of MZMs must be performed using \billchange{multiple wire sections}. As determined in Ref. \cite{Truong2023}, the total diabatic error from this transport is approximately the sum of the errors corresponding to each individual \billchange{section}---we expect that this behavior will persist when either disorder or noise is present. For coupling-based braiding schemes, we expect that the diabatic error will continue to be dominated by the low-energy dynamics, specifically the behavior of the minimum energy gap as each coupling is tuned.}

In summary, the results presented in this work elucidate the impact of disorder and noise on the operation of transporting MZMs and paints a clearer picture with regards to the underlying causes for increased diabatic errors. In view of applications toward braiding, possible future directions include the development of transport schemes which are capable of minimizing expected increases in the diabatic error. For example, it may be useful to study more elaborate protocols beyond the \billchange{usual piano key approach that this work is based on}. The characterization of the diabatic error in these settings remains an important subject for physical braiding of MZMs conducted on realistic experimental platforms.

\textit{Note added}. During the writing of this paper, we became aware of the recent independent work of Ref. \cite{Sahu2024}, which has a similar focus. 


\begin{acknowledgments}
BPT and TP acknowledge financial support from the Natural Sciences and Engineering Research Council of Canada (NSERC) and the Fonds de recherche du Qu\'{e}bec---Nature et technologies (FRQNT). BPT acknowledges support from the FRQNT doctoral training scholarships. KA acknowledges funding by US Department of Energy, Office of Science, Basic Energy Sciences, Materials Sciences and Engineering Division.
\end{acknowledgments}

\section*{Data availability}
The data that supports the findings of this article are openly available \cite{Truong2026_code}. 

\appendix
\section{Calculation of diabatic error using covariance matrix method} \label{app:covariance}
Following Refs. \cite{Bravyi2017, Bauer2018, Truong2023}, we provide an overview of the covariance matrix method applied to the calculation of the diabatic error. This overview is meant to be self-contained and only discusses details which are sufficient for the calculation. More extensive details may be found in Ref. \cite{Bravyi2017}.

It useful to first establish the relevant operators and matrix transformations using the Kitaev Hamiltonian. Up to a constant energy shift, the Kitaev Hamiltonian may be written as
\begin{equation}
	\billchange{H = \frac{1}{2} \mathbf{\Psi}^{\dagger} \mathcal{H} \mathbf{\Psi},} 
	\label{eq:app:hamkit}
\end{equation}
where $\mathbf{\Psi} = (c_{1}, c_{1}^{\dagger}, c_{2}, c_{2}^{\dagger}, ..., c_{N}, c_{N}^{\dagger})^{\mathrm{T}}$ is a vector of electron creation and annihilation operators and $\mathcal{H}$ is the single-particle Hamiltonian. We introduce the following Majorana operators in terms of the electron operators:
\begin{equation}
	\gamma_{2j-1} = c_{j} + c_{j}^{\dagger},~\gamma_{2j} = -i(c_{j} - c_{j}^{\dagger}),
	\label{eq:app:majops}
\end{equation}
such that they satisfy the properties $\gamma_{k} = \gamma_{k}^{\dagger}$, $\{\gamma_{k}, \gamma_{k'} \} = 2 \delta_{kk'}$. The Hamiltonian in Eq.~(\ref{eq:app:hamkit}) can be rewritten in terms of these Majorana operators as
\begin{equation}
	\billchange{H = \frac{i}{4} \mathbf{\Gamma}^{\mathrm{T}} \mathcal{A} \mathbf{\Gamma},}
	\label{eq:app:hamkit_maj}
\end{equation}
where $\mathbf{\Gamma} = (\gamma_{1}, \gamma_{2}, ..., \gamma_{2N})^{\mathrm{T}}$ and $\mathcal{A}$ is a real antisymmetric matrix. It is known that real antisymmetric matrices may be block-diagonalized as $\mathcal{A} = \mathcal{W}^{\mathrm{T}} \mathcal{B} \mathcal{W}$ using a real orthogonal transformation $\mathcal{W}$ with $\mathcal{B}$ taking the form:
\begin{equation}
	\mathcal{B} = \bigoplus_{j=1}^{N} 
	\begin{pmatrix}
		0 & \epsilon_{j}
		\\
		-\epsilon_{j} & 0
	\end{pmatrix},
	\label{eq:app:matB}
\end{equation}
where $\pm i\epsilon_{j}$ are the eigenvalues of $\mathcal{A}$ and $\epsilon_{j} > 0$ themselves may be interpreted as the non-negative single-particle energies. Inserting this block-diagonal transformation into Eq.~(\ref{eq:app:hamkit_maj}), we obtain 
\begin{equation}
	H = \frac{i}{4} \mathbf{\Xi}^{\mathrm{T}} \mathcal{B} \mathbf{\Xi} = \frac{i}{2} \sum_{j=1}^{N} \epsilon_{j} \eta_{2j-1} \eta_{2j},
	\label{eq:app:hamkit_maj_diag}
\end{equation}
where we have defined a vector $\mathbf{\Xi} \equiv \mathcal{W} \mathbf{\Gamma} = (\eta_{1}, \eta_{2}, ... , \eta_{2N})^{\mathrm{T}}$ which contains new Majorana operators $\eta_{k}$. Fermionic creation and annihilation operators corresponding to Bogoliubov quasiparticles may be constructed from these new Majorana operators as $d_{j} = (1/2)(\eta_{2j-1} + i \eta_{2j})$.

The covariance matrix $\mathcal{M}$ corresponding to the many-body ground state $| \Omega \rangle$ of the Kitaev Hamiltonian is formally defined as
\begin{equation}
	\mathcal{M}_{pq} = -\frac{i}{2} \langle \Omega |[\eta_{p}, \eta_{q}]| \Omega \rangle.
	\label{eq:app:covar_define}
\end{equation}
By invoking the Bogoliubov operators $d_{j}$, $d_{j}^{\dagger}$, one can show that the covariance matrix takes the form
\begin{equation}
	\mathcal{M} = \bigoplus_{j=1}^{N} i \sigma_{y},
	\label{eq:app:covar_define_bigo}
\end{equation}
where $\sigma_{y}$ is the second Pauli matrix. It is often practical to consider the covariance matrix in the basis of the original Majorana operators $\gamma_{k}$. Denoting this covariance matrix as $\mathcal{M}_{0}$, it is formally defined as
\begin{equation}
	\mathcal{M}_{0,pq} = -\frac{i}{2} \langle \Omega |[\gamma_{p}, \gamma_{q}]| \Omega \rangle.
	\label{eq:app:covar0_define}
\end{equation}
The change of basis is facilitated by using the following transformation:
\begin{equation}
	\mathcal{M}_{0} = \mathcal{W}^{\mathrm{T}} \mathcal{M} \mathcal{W},
	\label{eq:app:covar0_changebasis}
\end{equation}
where $\mathcal{W}$ is the same real orthogonal transformation previously shown to block-diagonalize the Hamiltonian in the original Majorana basis. The utility of performing this change of basis is illustrated when dynamics are considered. Suppose that these dynamics are captured by a single-particle time evolution operator $\mathcal{U}(\tau)$. The time evolution of the covariance matrix is then described by
\begin{equation}
	\mathcal{M}_{0}(\tau) = \mathcal{U}(\tau) \mathcal{M}_{0} \mathcal{U}(\tau)^{\dagger}.
	\label{eq:app:covar0_timeevolve}
\end{equation}

We are now equipped to construct the diabatic error in this formalism. Suppose that the covariance matrices corresponding to a final instantaneous ground state $| \Omega_{\mathrm{f}} \rangle$ and a time-evolved ground state $ U(\tau)| \Omega_{\mathrm{i}} \rangle$ are denoted as $\mathcal{M}_{0,\mathrm{f}}$ and $\mathcal{M}_{0,\mathrm{i}}(\tau)$, respectively. Using the overlap formula provided in Ref. \cite{Bravyi2017}, the diabatic error may be stated as
\begin{equation}
	\mathcal{P} = 1 - 2^{-N} |\mathrm{Pf} (\mathcal{M}_{0,\mathrm{f}} + \mathcal{M}_{0,\mathrm{i}}(\tau))|,
	\label{eq:app:diaberr_covars}
\end{equation}
where $\mathrm{Pf}( \cdots )$ is the Pfaffian. In our numerics, we use the methods of Ref. \cite{Wimmer2012} to calculate the Pfaffian as well as the orthogonal transformations $\mathcal{W}$.

\section{Time evolution operator} \label{app:timeevol}
Following Refs. \cite{Bauer2018, Truong2023}, we briefly detail the calculation of the time evolution operator which encodes the transport protocol. To conform with the covariance matrix method described in Appendix \ref{app:covariance}, the single-particle time evolution operator $\mathcal{U}(\tau)$ must be represented in the basis of the original Majorana operators $\gamma_{k}$. This can be done by considering the matrix $\mathcal{A}(t)$, where we have explicitly introduced time-dependence. In general, the single-particle time evolution operator is
\begin{equation}
	\mathcal{U}(\tau) = \mathcal{T} \exp \left( \int_{0}^{\tau} dt \mathcal{A}(t) \right),
	\label{eq:app:timeevol_gen}
\end{equation}
where $\mathcal{T}$ denotes time ordering. In our numerics, Eq.~(\ref{eq:app:timeevol_gen}) is approximated by discretizing time which results in the following time-ordered product of matrix exponentials:
\begin{equation}
	\mathcal{U}(\tau) \approx \mathcal{T} \prod_{p=1}^{N_{\mathrm{s}}} e^{\Delta t \mathcal{A}(t_{p})},
	\label{eq:app:timeevol_prod}
\end{equation}
where $\Delta t = t_{p+1} - t_{p}$ is the time step and $N_{\mathrm{s}} = \tau/ \Delta t$ is the number of time steps. In our simulations, we use $\Delta t \sim 10^{-2} ~\text{meV}^{-1}$ and $N_{\mathrm{s}}$ ranges from $10^{3}$--$10^{4}$. For reference, the typical time-scales that we consider in our simulations are $\tau \sim \tau_{\mathrm{LZ}} \sim 10^{2} ~\text{meV}^{-1}$. All times have dimensions of inverse energy since we set $\hbar = 1$.

\section{Application of perturbation theory to minimum gap statistics} \label{app:ptheory}
\begin{figure*}[t]
	\centering
	\phantomsubfloat{\label{fig:uncorr_mingap_expr_a}}
	\phantomsubfloat{\label{fig:uncorr_mingap_expr_b}}
	\includegraphics[width=1.0\textwidth]{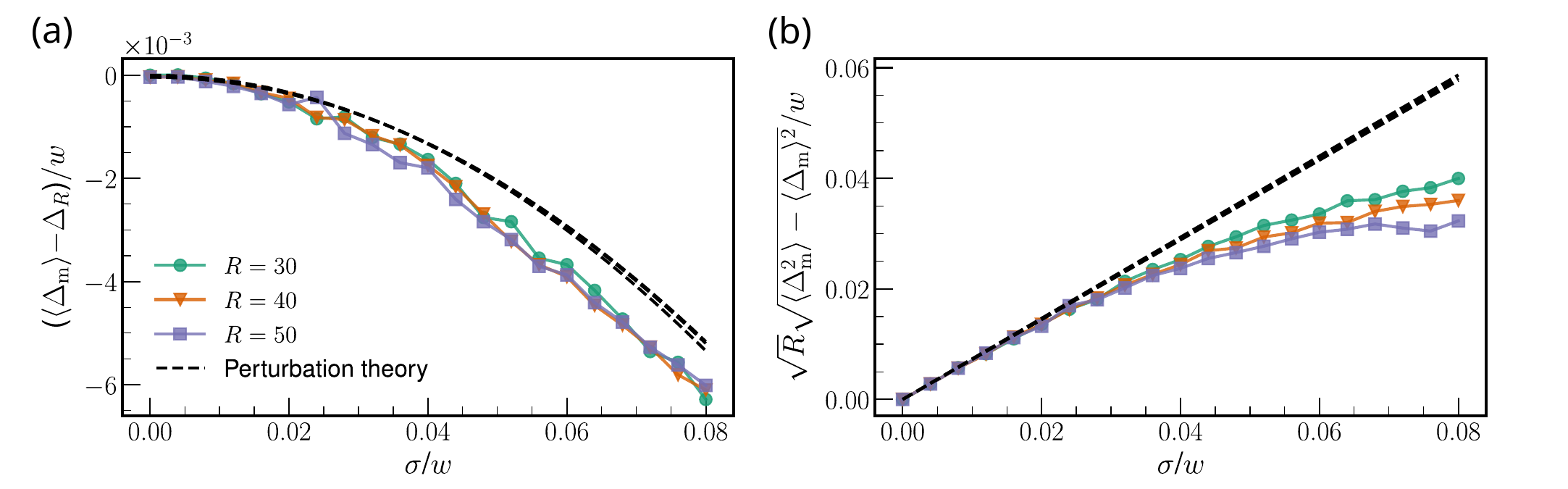}
    \caption[]{Numerical results for the minimum gap statistics for uncorrelated disorder in Fig.~\ref{fig:uncorr_mingap} in comparison to the semianalytical expressions of the perturbation theory approach. (a) Average minimum gap and perturbation theory expression (dashed black lines) calculated semianalytically from Eqs.~(\ref{eq:app:mingap_perturb_avg}) and (\ref{eq:app:secondcorr_avg}). (b) Standard deviation and perturbation theory expression (dashed black lines) calculated semianalytically from Eqs.~(\ref{eq:app:mingap_perturb_std}) and (\ref{eq:app:firstcorr_sqr_avg}).}
  	\label{fig:app:uncorr_mingap_expr}
\end{figure*}
\begin{figure*}[t]
	\centering
	\includegraphics[width=1.0\textwidth]{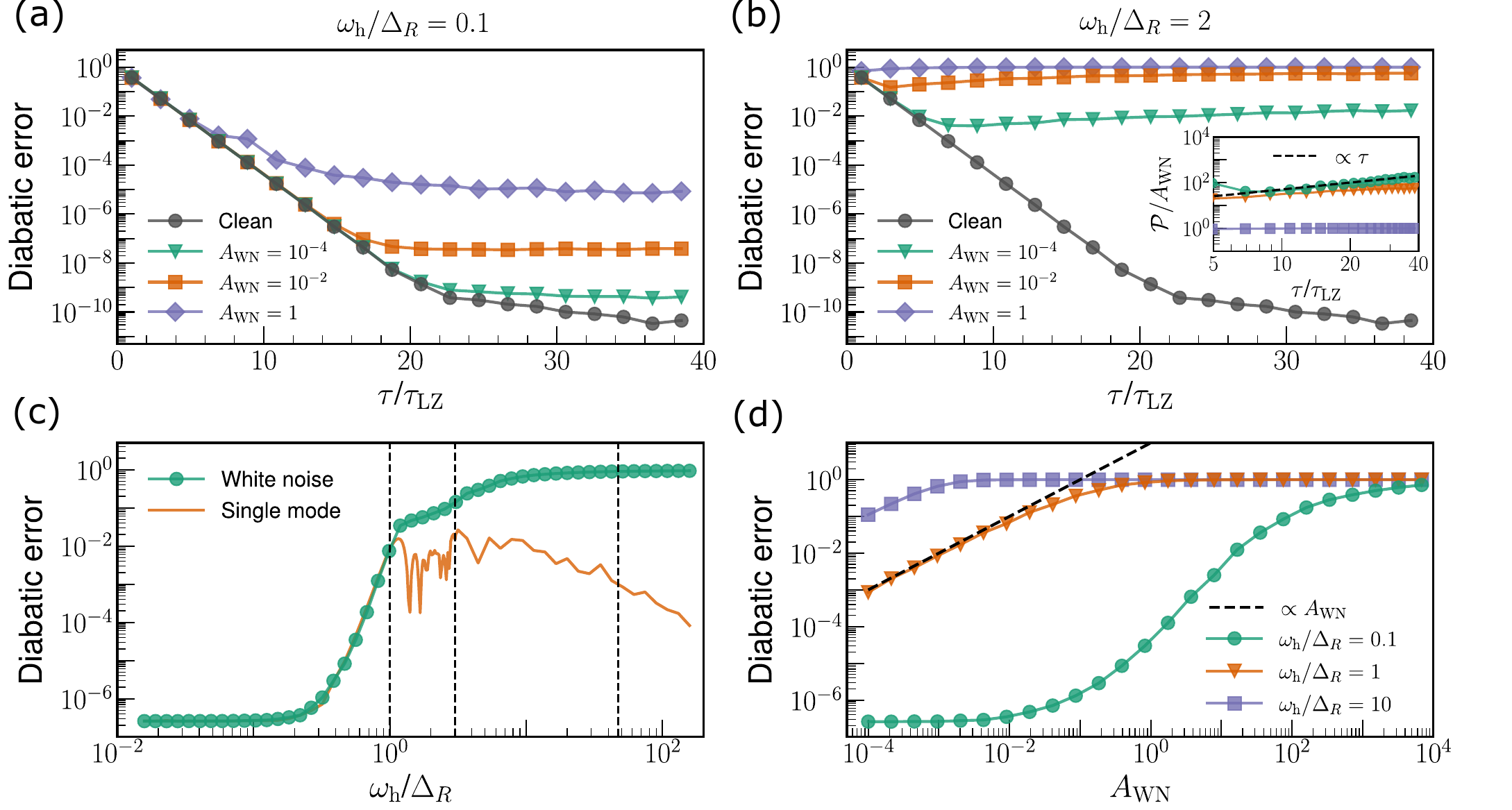}
    \caption[]{Numerical results for the average diabatic error at the end of the transport protocol simulated with white noise added to the chemical potential tuning. (a) Diabatic error versus protocol time for select noise amplitudes \billchange{(in units of meV)} with high-frequency cutoff $\omega_{\mathrm{h}}/\Delta_{R} = 0.1$ which is below the minimum gap. (b) Same as panel (a) but with high-frequency cutoff $\omega_{\mathrm{h}}/\Delta_{R} = 2$ which is above the minimum gap. The inset illustrates the linear rise in the average diabatic error for sufficiently long protocol times. (c) Diabatic error versus high-frequency cutoff in the case of white noise and frequency in the case of a single mode with protocol time fixed to $\tau/\tau_{\mathrm{LZ}} = 15$ for both. The white noise amplitude is \billchange{$A_{\mathrm{WN}} = 10^{-3}$ meV} while the single mode amplitude is \billchange{$A_{\mathrm{SM}} = 5 \times 10^{-3}$ meV}. Here, in order from left to right, the dashed black lines correspond to the clean minimum gap, the clean maximum gap, and the hopping amplitude. (d) Diabatic error versus noise amplitude for select high-frequency cutoffs with protocol time $\tau/\tau_{\mathrm{LZ}} = 15$. Unless otherwise specified, the default chain parameters are identical to those cited in Fig.~\ref{fig:uncorr_diab} and in the main text. Averaging is performed over $500$ simulations for each result. For the single-mode results, averaging is performed over the phase.}
  	\label{fig:app:whitenoise}
\end{figure*}
We demonstrate that perturbation theory can be used to adequately describe the minimum gap statistics in the case of weak uncorrelated disorder. To begin, we recall the definition of the minimum gap provided in the main text:
\begin{equation}
	\Delta_{\mathrm{m}} = \mathrm{min}\{\epsilon_{1}(t) - \epsilon_{0}(t)\},
	\label{eq:app:mingap}
\end{equation}
where $\epsilon_{n}$ is the $n$th nonnegative single-particle energy. Before proceeding with further calculations, we make two simplifying assumptions. First, the energy $\epsilon_{0}(t)$, which is always the MZM energy in our system, is assumed to be much smaller compared to the energy of the lowest energy bulk mode $\epsilon_{1}(t)$, i.e. $\epsilon_{0}(t) \ll \epsilon_{1}(t)$. Second, the minimum gap is assumed to occur when the \billchange{right section's} chemical potential achieves criticality, i.e. $\mu(\tau/2) = w$. Treating the disorder strength $\sigma = \sqrt{ \llangle \Delta_{\mathrm{m}}^2 \rrangle - \llangle \Delta_{\mathrm{m}} \rrangle^2}$ as a small parameter compared to the clean minimum gap, we write the minimum gap in terms of correction up to second order:
\begin{equation}
	\Delta_{\mathrm{m}} = \epsilon_{1}^{(0)} + \epsilon_{1}^{(1)} + \epsilon_{1}^{(2)},
	\label{eq:app:mingap_perturb}
\end{equation}
where $\epsilon_{1}^{(k)}$ is the $k$th-order correction to $\epsilon_{1}(\tau/2)$. The zeroth-order term roughly corresponds to the clean minimum gap, $\epsilon_{1}^{(0)} \approx \Delta_{R} = \pi \Delta_{\mathrm{SC}}/R$. Performing an average over different minimum gap realizations results in the average
\begin{equation}
	\llangle \Delta_{\mathrm{m}} \rrangle = \epsilon_{1}^{(0)} + \llangle \epsilon_{1}^{(1)} \rrangle + \llangle \epsilon_{1}^{(2)} \rrangle,
	\label{eq:app:mingap_perturb_avg}
\end{equation}
as well as the variance
\begin{equation}
	\sigma^2_{\mathrm{m}} \equiv \llangle \Delta^2_{\mathrm{m}} \rrangle - \llangle \Delta_{\mathrm{m}} \rrangle^2 = \llangle ( \epsilon_{1}^{(1)} )^2 \rrangle.
	\label{eq:app:mingap_perturb_std}
\end{equation}

We now calculate the first and second-order corrections, $\epsilon_{1}^{(1)}$ and $\epsilon_{1}^{(2)}$, respectively, while regarding the disorder as a static perturbation. To set up this calculation, we establish the Hamiltonian explicitly in terms of its unperturbed and perturbed components:
\begin{equation}
	H = H_{0} + V,
	\label{eq:app:H}
\end{equation}
where $H_{0}$ corresponds to a clean Kitaev chain with the \billchange{right section} at criticality:
\begin{align}
	H_{0} = &-\mu_{\mathrm{L}} \sum_{j=1}^{L} c_{j}^{\dagger} c_{j} - w \sum_{j=L+1}^{N} c_{j}^{\dagger} c_{j} 
	\nonumber	
	\\
	&- \frac{w}{2} \sum_{j=1}^{N-1} (c_{j}^{\dagger} c_{j+1} + \text{H.c.}) -\frac{\Delta_{\mathrm{SC}}}{2} \sum_{j=1}^{N-1} (c_{j} c_{j+1} + \text{H.c.}), 
	\label{eq:app:H0_kitaev}
\end{align}
and $V$ corresponds to the disorder potential:
\begin{equation}
	V = -\sum_{j = 1}^{N} \delta \mu_{j} c_{j}^{\dagger} c_{j}.
	\label{eq:app:V_disorder}
\end{equation}

Referring to Eq.~(\ref{eq:app:hamkit}) and the accompanying discussion in Appendix \ref{app:covariance}, suppose that we denote $\mathcal{H}_{0}$ and $\mathcal{V}$ as the single-particle versions of Eqs.~(\ref{eq:app:H0_kitaev}) and (\ref{eq:app:V_disorder}), respectively. The eigenvectors of $\mathcal{H}_{0}$ are used to calculate the correction terms we desire. The eigenvectors corresponding to positive energies $\epsilon_{n}$ are denoted as 
\begin{equation}
	\vec{\psi}^{\mathrm{T}}_{n} = (u_{1}^{(n)}, v_{1}^{(n)}, u_{2}^{(n)}, v_{2}^{(n)}, ..., u_{N}^{(n)}, v_{N}^{(n)})^{\mathrm{T}},
	\label{eq:app:eigvecs}
\end{equation}
for complex coherence factors $u_{j}^{(n)}$ and $v_{j}^{(n)}$. The eigenvectors $\vec{\phi}_{n}$ corresponding to negative energies $-\epsilon_{n}$ may be obtained by performing a particle-hole transformation on Eq.~(\ref{eq:app:eigvecs}) such that $u_{j}^{(n)} \rightarrow v_{j}^{(n)*}$ and $v_{j}^{(n)} \rightarrow -u_{j}^{(n)*}$. The first-order correction to the minimum gap is calculated using the $n = 1$ eigenvector which corresponds to the lowest energy bulk mode:
\begin{equation}
	\epsilon_{1}^{(1)} = \vec{\psi}^{\dagger}_{1} \mathcal{V} \vec{\psi}_{1} = -\sum_{j=1}^{N} \delta \mu_{j} \left( |u_{j}^{(1)}|^2 - |v_{j}^{(1)}|^2 \right).
	\label{eq:app:firstcorr}
\end{equation}
It is immediately clear that Eq.~(\ref{eq:app:firstcorr}) averages to zero since $\llangle \delta \mu_{j} \rrangle = 0$ by construction. This suggests that the average minimum gap given in Eq.~(\ref{eq:app:mingap_perturb_avg}) depends chiefly on the second-order correction. However, the average of the square of Eq.~(\ref{eq:app:firstcorr}), which enters the variance in Eq.~(\ref{eq:app:mingap_perturb_std}), is generally nonzero:
\begin{equation}
	\llangle (\epsilon_{1}^{(1)})^2 \rrangle = \sigma^2 \sum_{j=1}^{N} \left( |u_{j}^{(1)}|^2 - |v_{j}^{(1)}|^2 \right)^2,
	\label{eq:app:firstcorr_sqr_avg}
\end{equation}
where the variance $\sigma^2$ of the disorder potential is defined by $\llangle \delta \mu_{j} \delta \mu_{j'} \rrangle = \delta_{jj'} \sigma^2$. The second-order correction, which enters the average minimum gap in Eq.~(\ref{eq:app:mingap_perturb_avg}), is given by 
\begin{align}
	\epsilon_{1}^{(2)} &= \sum_{n>1}^{N} \left( \frac{ |\vec{\psi}^{\dagger}_{1} \mathcal{V} \vec{\psi}_{n}|^2 }{\epsilon_{1}^{(0)} - \epsilon_{n}^{(0)}} + \frac{ |\vec{\psi}^{\dagger}_{1} \mathcal{V} \vec{\phi}_{n}|^2 }{\epsilon_{1}^{(0)} + \epsilon_{n}^{(0)}} \right)
	\nonumber	
	\\[10pt]
	&= \sum_{n>1}^{N} \left[ \frac{ \left| \sum_{j=1}^{N} \delta \mu_{j} \left( u_{j}^{(1)} u_{j}^{(n)*} - v_{j}^{(1)} v_{j}^{(n)*} \right) \right|^2 }{\epsilon_{1}^{(0)} - \epsilon_{n}^{(0)}} \right.
	\nonumber
	\\[10pt]	
	&+ \left. \frac{ \left| \sum_{j=1}^{N} \delta \mu_{j} \left( u_{j}^{(1)} v_{j}^{(n)*} - v_{j}^{(1)} u_{j}^{(n)*} \right) \right|^2 }{\epsilon_{1}^{(0)} + \epsilon_{n}^{(0)}} \right].
	\label{eq:app:secondcorr}
\end{align}
Upon averaging, Eq.~(\ref{eq:app:secondcorr}) becomes
\begin{align}
	\llangle \epsilon_{1}^{(2)} \rrangle &= \sigma^2 \sum_{n>1}^{N} \left[ \frac{ \sum_{j=1}^{N} \left| u_{j}^{(1)} u_{j}^{(n)*} - v_{j}^{(1)} v_{j}^{(n)*} \right|^2 }{\epsilon_{1}^{(0)} - \epsilon_{n}^{(0)}} \right.
	\nonumber
	\\[10pt]	
	&+ \left. \frac{ \sum_{j=1}^{N} \left| u_{j}^{(1)} v_{j}^{(n)*} - v_{j}^{(1)} u_{j}^{(n)*} \right|^2 }{\epsilon_{1}^{(0)} + \epsilon_{n}^{(0)}} \right].
	\label{eq:app:secondcorr_avg}
\end{align}

A comparison between this perturbation theory approach and the minimum gap statistics for a transport protocol with uncorrelated disorder is shown in Fig.~\ref{fig:app:uncorr_mingap_expr}. The expressions for the average and standard deviation are calculated from Eqs.~(\ref{eq:app:mingap_perturb_avg}) and (\ref{eq:app:mingap_perturb_std}) seminumerically, meaning that the coherence factors and the energies are obtained by numerically diagonalizing the Hamiltonian of the clean Kitaev chain with the \billchange{right section} at criticality. We see that for weak disorder, the perturbation theory approach largely agrees with the numerics. For the average minimum gap in particular, the perturbation theory approach continues to show some agreement even at larger disorder strengths. Importantly, the observed scaling of the numerical results with the \billchange{transport distance} $R$ is respected by the perturbation theory approach. 

The calculations presented in this Appendix are observed to be suitable for characterizing the minimum gap statistics for weak uncorrelated disorder. For the case of correlated disorder, additional complications arise which conflict with the initial assumptions pertaining to our approach. Specifically, we had assumed that the average minimum gap occurs roughly when $\mu(\tau/2) = w$, which coincides with the \billchange{right wire section being} at criticality in a clean chain. While this may be approximately true for uncorrelated disorder, we verify that this assumption fails when sufficiently strong correlations are present. We speculate that the behavior of the minimum gap for correlated disorder is susceptible to the presence of localized bound states, which are more likely to emerge in this setting given the ability of the disorder to induce isolated potential wells and barriers. These bound states may be capable of suppressing the minimum gap further away from the clean criticality condition. \billchange{Importantly, perturbation theory is not expected to be capable of finding these bound states because they emerge non-perturbatively and so their effects on the minimum gap cannot be adequately captured using this approach.}

\section{White noise} \label{app:whitenoise}
We demonstrate the effect of white noise on the diabatic error of the transport protocol. The results shown in this Appendix serve to complement those corresponding to $1/f$ noise in Sec.~\ref{sec:1fnoise}. For each simulation of the transport protocol, we generate a white noise signal with cutoff frequencies in accordance with the procedure outlined in Sec.~\ref{sec:prelim:noise}. The power spectral density is
\begin{equation}
	S(\omega_{k}) = 
	\begin{cases}
		A_{\mathrm{WN}}, &\text{for}~ \omega_{\mathrm{l}} \leq \omega_{k} \leq \omega_{\mathrm{h}},
		\\
		0, &\text{otherwise},
	\end{cases}
\end{equation}
where $\omega_{\mathrm{l}}$ and $\omega_{\mathrm{h}}$ remain as the low and high-frequency cutoffs, respectively. The transport protocol is simulated with white noise acting only on the \billchange{right section's} chemical potential tuning. Our results are illustrated in Fig.~\ref{fig:app:whitenoise}. Here, each diabatic error is calculated and averaged over 500 simulations which differ only in their noise signal realization. Only the standard average is shown since its difference with the geometric average is found to be relatively small. The default chain parameters are identical to those used in the main text unless otherwise stated. Qualitatively, the observations discussed in Sec.~\ref{sec:1fnoise} for $1/f$ noise are completely applicable to white noise. The diabatic error still features two distinct long-time behaviors depending on the cutoff frequencies. As well, for high-frequency cutoffs beyond the clean minimum gap, the error suffers a drastic increase, indicating a decreased level of success for the protocol in this regime. 

\clearpage
\newpage

\bibliography{references.bib}

@article{Truong2026_code,
  title = {{Numerical code and data for "Shuttling Majorana zero modes in disordered and noisy topological superconductors"}},
  author = {Truong, Bill and Agarwal, Kartiek and Pereg-Barnea, Tami},
  journal = {Zenodo},
  month = {Jan},
  year = {2026},
  publisher = {Zenodo},
  doi = {10.5281/zenodo.18396960},
  url = {https://doi.org/10.5281/zenodo.18396960},
}

@article{mori2016rigorous,
  title = {{Rigorous Bound on Energy Absorption and Generic Relaxation in Periodically Driven Quantum Systems}},
  author = {Mori, Takashi and Kuwahara, Tomotaka and Saito, Keiji},
  journal = {Phys. Rev. Lett.},
  volume = {116},
  issue = {12},
  pages = {120401},
  numpages = {5},
  year = {2016},
  month = {Mar},
  publisher = {American Physical Society},
  doi = {10.1103/PhysRevLett.116.120401},
  url = {https://link.aps.org/doi/10.1103/PhysRevLett.116.120401}
}

@article{agarwal2020dynamical,
  title = {{Dynamical Enhancement of Symmetries in Many-Body Systems}},
  author = {Agarwal, Kartiek and Martin, Ivar},
  journal = {Phys. Rev. Lett.},
  volume = {125},
  issue = {8},
  pages = {080602},
  numpages = {6},
  year = {2020},
  month = {Aug},
  publisher = {American Physical Society},
  doi = {10.1103/PhysRevLett.125.080602},
  url = {https://link.aps.org/doi/10.1103/PhysRevLett.125.080602}
}

@article{martin2022effect,
  title = {Effect of quasiperiodic and random noise on many-body dynamical decoupling protocols},
  author = {Martin, Tristan and Martin, Ivar and Agarwal, Kartiek},
  journal = {Phys. Rev. B},
  volume = {106},
  issue = {13},
  pages = {134306},
  numpages = {15},
  year = {2022},
  month = {Oct},
  publisher = {American Physical Society},
  doi = {10.1103/PhysRevB.106.134306},
  url = {https://link.aps.org/doi/10.1103/PhysRevB.106.134306}
}

@article{min2022dynamical,
  title = {Dynamical approach to improving Majorana qubits and distinguishing them from trivial bound states},
  author = {Min, Brett and Fajardo, Bastien and Pereg-Barnea, T. and Agarwal, Kartiek},
  journal = {Phys. Rev. B},
  volume = {105},
  issue = {15},
  pages = {155412},
  numpages = {20},
  year = {2022},
  month = {Apr},
  publisher = {American Physical Society},
  doi = {10.1103/PhysRevB.105.155412},
  url = {https://link.aps.org/doi/10.1103/PhysRevB.105.155412}
}

@article{Read2000,
  title = {Paired states of fermions in two dimensions with breaking of parity and time-reversal symmetries and the fractional quantum Hall effect},
  author = {Read, N. and Green, Dmitry},
  journal = {Phys. Rev. B},
  volume = {61},
  issue = {15},
  pages = {10267--10297},
  numpages = {0},
  year = {2000},
  month = {Apr},
  publisher = {American Physical Society},
  doi = {10.1103/PhysRevB.61.10267},
  url = {https://link.aps.org/doi/10.1103/PhysRevB.61.10267}
}

@article{Kitaev2001,
   author = {A Yu Kitaev},
   doi = {10.1070/1063-7869/44/10S/S29},
   issue = {131},
   journal = {Phys. Usp.},
   pages = {131},
   publisher = {Plenum Press},
   title = {Unpaired {M}ajorana fermions in quantum wires},
   volume = {44},
   url = {https://doi.org/10.1070/1063-7869/44/10S/S29},
   year = {2001},
}

@article{Ivanov2001,
   author = {D. A. Ivanov},
   doi = {10.1103/PhysRevLett.86.268},
   issn = {00319007},
   issue = {2},
   journal = {Phys. Rev. Lett.},
   month = {1},
   pages = {268-271},
   publisher = {American Inst of Physics},
   title = {Non-{A}belian statistics of half-quantum vortices in $p$-wave superconductors},
   volume = {86},
   url = {https://doi.org/10.1103/PhysRevLett.86.268},
   year = {2001},
}

@article{Kitaev2003,
   author = {A Yu Kitaev},
   doi = {10.1016/S0003-4916(02)00018-0},
   issue = {1},
   journal = {Ann. Phys.},
   pages = {2-30},
   title = {Fault-tolerant quantum computation by anyons},
   volume = {303},
   url = {https://doi.org/10.1016/S0003-4916(02)00018-0},
   year = {2003},
}

@article{Fu2008,
  title = {{Superconducting Proximity Effect and Majorana Fermions at the Surface of a Topological Insulator}},
  author = {Fu, Liang and Kane, C. L.},
  journal = {Phys. Rev. Lett.},
  volume = {100},
  issue = {9},
  pages = {096407},
  numpages = {4},
  year = {2008},
  month = {Mar},
  publisher = {American Physical Society},
  doi = {10.1103/PhysRevLett.100.096407},
  url = {https://link.aps.org/doi/10.1103/PhysRevLett.100.096407}
}

@article{Fu2009,
  title = {Josephson current and noise at a superconductor/quantum-spin-Hall-insulator/superconductor junction},
  author = {Fu, Liang and Kane, C. L.},
  journal = {Phys. Rev. B},
  volume = {79},
  issue = {16},
  pages = {161408},
  numpages = {4},
  year = {2009},
  month = {Apr},
  publisher = {American Physical Society},
  doi = {10.1103/PhysRevB.79.161408},
  url = {https://link.aps.org/doi/10.1103/PhysRevB.79.161408}
}

@article{Nayak2008,
   author = {Nayak, Chetan and Simon, Steven H. and Stern, Ady and Freedman, Michael and Das Sarma, Sankar},
   doi = {10.1103/RevModPhys.80.1083},
   issn = {00346861},
   issue = {3},
   journal = {Rev. Mod. Phys.},
   month = {9},
   pages = {1083-1159},
   title = {Non-{A}belian anyons and topological quantum computation},
   volume = {80},
   url = {https://doi.org/10.1103/RevModPhys.80.1083},
   year = {2008},
}

@article{DasSarma2015,
   author = {Sankar Das Sarma and Michael Freedman and Chetan Nayak},
   doi = {10.1038/NPJQI.2015.1},
   issn = {20566387},
   journal = {npj Quantum Inf.},
   month = {12},
   pages = {15001},
   publisher = {Nature Research},
   title = {Majorana zero modes and topological quantum computation},
   volume = {1},
   url = {https://doi.org/10.1038/NPJQI.2015.1},
   year = {2015},
}

@article{Lutchyn2010,
  title = {{Majorana Fermions and a Topological Phase Transition in Semiconductor-Superconductor Heterostructures}},
  author = {Lutchyn, Roman M. and Sau, Jay D. and Das Sarma, S.},
  journal = {Phys. Rev. Lett.},
  volume = {105},
  issue = {7},
  pages = {077001},
  numpages = {4},
  year = {2010},
  month = {Aug},
  publisher = {American Physical Society},
  doi = {10.1103/PhysRevLett.105.077001},
  url = {https://link.aps.org/doi/10.1103/PhysRevLett.105.077001}
}

@article{Oreg2010,
  title = {{Helical Liquids and Majorana Bound States in Quantum Wires}},
  author = {Oreg, Yuval and Refael, Gil and von Oppen, Felix},
  journal = {Phys. Rev. Lett.},
  volume = {105},
  issue = {17},
  pages = {177002},
  numpages = {4},
  year = {2010},
  month = {Oct},
  publisher = {American Physical Society},
  doi = {10.1103/PhysRevLett.105.177002},
  url = {https://link.aps.org/doi/10.1103/PhysRevLett.105.177002}
}

@article{Sau2010,
  title = {Non-{Abelian} quantum order in spin-orbit-coupled semiconductors: {Search} for topological {Majorana} particles in solid-state systems},
  author = {Sau, Jay D. and Tewari, Sumanta and Lutchyn, Roman M. and Stanescu, Tudor D. and Das Sarma, S.},
  journal = {Phys. Rev. B},
  volume = {82},
  issue = {21},
  pages = {214509},
  numpages = {26},
  year = {2010},
  month = {Dec},
  publisher = {American Physical Society},
  doi = {10.1103/PhysRevB.82.214509},
  url = {https://link.aps.org/doi/10.1103/PhysRevB.82.214509}
}

@article{Sau2010a,
  title = {Generic {{New Platform}} for {{Topological Quantum Computation Using Semiconductor Heterostructures}}},
  author = {Sau, Jay D. and Lutchyn, Roman M. and Tewari, Sumanta and Das Sarma, S.},
  year = {2010},
  month = jan,
  journal = {Phys. Rev. Lett.},
  volume = {104},
  number = {4},
  pages = {040502},
  issn = {0031-9007, 1079-7114},
  doi = {10.1103/PhysRevLett.104.040502},
  url = {https://link.aps.org/doi/10.1103/PhysRevLett.104.040502}
}

@article{Alicea2010,
  title = {Majorana Fermions in a Tunable Semiconductor Device},
  author = {Alicea, Jason},
  year = {2010},
  month = mar,
  journal = {Phys. Rev. B},
  volume = {81},
  number = {12},
  pages = {125318},
  issn = {1098-0121, 1550-235X},
  doi = {10.1103/PhysRevB.81.125318},
  url = {https://link.aps.org/doi/10.1103/PhysRevB.81.125318}
}

@article{Stanescu2011,
  title = {Majorana fermions in semiconductor nanowires},
  author = {Stanescu, Tudor D. and Lutchyn, Roman M. and Das Sarma, S.},
  journal = {Phys. Rev. B},
  volume = {84},
  issue = {14},
  pages = {144522},
  numpages = {29},
  year = {2011},
  month = {Oct},
  publisher = {American Physical Society},
  doi = {10.1103/PhysRevB.84.144522},
  url = {https://link.aps.org/doi/10.1103/PhysRevB.84.144522}
}

@article{Cook2011,
  title = {Majorana fermions in a topological-insulator nanowire proximity-coupled to an $s$-wave superconductor},
  author = {Cook, A. and Franz, M.},
  journal = {Phys. Rev. B},
  volume = {84},
  issue = {20},
  pages = {201105},
  numpages = {4},
  year = {2011},
  month = {Nov},
  publisher = {American Physical Society},
  doi = {10.1103/PhysRevB.84.201105},
  url = {https://link.aps.org/doi/10.1103/PhysRevB.84.201105}
}

@article{Hell2017,
  title = {Two-{{Dimensional Platform}} for {{Networks}} of {{Majorana Bound States}}},
  author = {Hell, Michael and Leijnse, Martin and Flensberg, Karsten},
  year = {2017},
  month = mar,
  journal = {Phys. Rev. Lett.},
  volume = {118},
  number = {10},
  pages = {107701},
  issn = {0031-9007, 1079-7114},
  doi = {10.1103/PhysRevLett.118.107701},
  url = {https://link.aps.org/doi/10.1103/PhysRevLett.118.107701}
}

@article{Pientka2017,
  title = {Topological {{Superconductivity}} in a {{Planar Josephson Junction}}},
  author = {Pientka, Falko and Keselman, Anna and Berg, Erez and Yacoby, Amir and Stern, Ady and Halperin, Bertrand I.},
  year = {2017},
  month = may,
  journal = {Phys. Rev. X},
  volume = {7},
  number = {2},
  pages = {021032},
  issn = {2160-3308},
  doi = {10.1103/PhysRevX.7.021032},
  url = {http://link.aps.org/doi/10.1103/PhysRevX.7.021032}
}

@article{Hegde2020,
  title = {A Topological {{Josephson}} Junction Platform for Creating, Manipulating, and Braiding {{Majorana}} Bound States},
  author = {Hegde, Suraj S. and Yue, Guang and Wang, Yuxuan and Huemiller, Erik and Van Harlingen, D.J. and Vishveshwara, Smitha},
  year = {2020},
  month = dec,
  journal = {Annals of Physics},
  volume = {423},
  pages = {168326},
  issn = {00034916},
  doi = {10.1016/j.aop.2020.168326},
  url = {https://linkinghub.elsevier.com/retrieve/pii/S0003491620302608}
}

@article{Choy2011,
  title = {Majorana fermions emerging from magnetic nanoparticles on a superconductor without spin-orbit coupling},
  author = {Choy, T.-P. and Edge, J. M. and Akhmerov, A. R. and Beenakker, C. W. J.},
  journal = {Phys. Rev. B},
  volume = {84},
  issue = {19},
  pages = {195442},
  numpages = {6},
  year = {2011},
  month = {Nov},
  publisher = {American Physical Society},
  doi = {10.1103/PhysRevB.84.195442},
  url = {https://link.aps.org/doi/10.1103/PhysRevB.84.195442}
}

@article{Nadj-Perge2013,
  title = {Proposal for realizing {Majorana} fermions in chains of magnetic atoms on a superconductor},
  author = {Nadj-Perge, S. and Drozdov, I. K. and Bernevig, B. A. and Yazdani, Ali},
  journal = {Phys. Rev. B},
  volume = {88},
  issue = {2},
  pages = {020407},
  numpages = {5},
  year = {2013},
  month = {Jul},
  publisher = {American Physical Society},
  doi = {10.1103/PhysRevB.88.020407},
  url = {https://link.aps.org/doi/10.1103/PhysRevB.88.020407}
}

@article{Braunecker2013,
  title = {{Interplay between Classical Magnetic Moments and Superconductivity in Quantum One-Dimensional Conductors: Toward a Self-Sustained Topological Majorana Phase}},
  author = {Braunecker, Bernd and Simon, Pascal},
  journal = {Phys. Rev. Lett.},
  volume = {111},
  issue = {14},
  pages = {147202},
  numpages = {5},
  year = {2013},
  month = {Oct},
  publisher = {American Physical Society},
  doi = {10.1103/PhysRevLett.111.147202},
  url = {https://link.aps.org/doi/10.1103/PhysRevLett.111.147202}
}

@article{Klinovaja2013,
  title = {Topological {{Superconductivity}} and {{Majorana Fermions}} in {{RKKY Systems}}},
  author = {Klinovaja, Jelena and Stano, Peter and Yazdani, Ali and Loss, Daniel},
  year = {2013},
  month = nov,
  journal = {Phys. Rev. Lett.},
  volume = {111},
  number = {18},
  pages = {186805},
  issn = {0031-9007, 1079-7114},
  doi = {10.1103/PhysRevLett.111.186805},
  url = {https://link.aps.org/doi/10.1103/PhysRevLett.111.186805}
}

@article{Pientka2013,
  title = {Topological superconducting phase in helical {Shiba} chains},
  author = {Pientka, Falko and Glazman, Leonid I. and von Oppen, Felix},
  journal = {Phys. Rev. B},
  volume = {88},
  issue = {15},
  pages = {155420},
  numpages = {13},
  year = {2013},
  month = {Oct},
  publisher = {American Physical Society},
  doi = {10.1103/PhysRevB.88.155420},
  url = {https://link.aps.org/doi/10.1103/PhysRevB.88.155420}
}

@article{Leijnse2012,
  title = {Parity qubits and poor man's {Majorana} bound states in double quantum dots},
  author = {Leijnse, Martin and Flensberg, Karsten},
  journal = {Phys. Rev. B},
  volume = {86},
  issue = {13},
  pages = {134528},
  numpages = {7},
  year = {2012},
  month = {Oct},
  publisher = {American Physical Society},
  doi = {10.1103/PhysRevB.86.134528},
  url = {https://link.aps.org/doi/10.1103/PhysRevB.86.134528}
}

@article{Sau2012,
   author = {Sau, J. D. and Sarma, S. Das},
   title = {Realizing a robust practical {Majorana} chain in a quantum-dot-superconductor linear array},
   doi = {10.1038/ncomms1966},
   journal = {Nat. Commun.},
   volume = {3},
   pages = {964},
   url = {https://doi.org/10.1038/ncomms1966},
   year = {2012},
}

@article{Fulga2013,
doi = {10.1088/1367-2630/15/4/045020},
url = {https://dx.doi.org/10.1088/1367-2630/15/4/045020},
year = {2013},
month = {apr},
publisher = {IOP Publishing},
volume = {15},
number = {4},
pages = {045020},
author = {Fulga, Ion C and Haim, Arbel and Akhmerov, Anton R and Oreg, Yuval},
title = {Adaptive tuning of {Majorana} fermions in a quantum dot chain},
journal = {New J. Phys.},
}

@article{Lutchyn2018,
   author = {Lutchyn, R. M. and P. A. M. Bakkers, E. and P. Kouwenhoven, L and Krogstrup, P and M. Marcus, C and Oreg, Y.},
   title = {Majorana zero modes in superconductor–semiconductor heterostructures},
   doi = {10.1038/s41578-018-0003-1},
   journal = {Nat. Rev. Mater.},
   volume = {3},
   pages = {52–68},
   url = {https://doi.org/10.1038/s41578-018-0003-1},
   year = {2018},
}

@article{Flensberg2021,
   author = {Flensberg, K and von Oppen, Felix and Stern, A},
   title = {Engineered platforms for topological superconductivity and {Majorana} zero modes},
   doi = {10.1038/s41578-021-00336-6},
   journal = {Nat. Rev. Mater.},
   volume = {6},
   pages = {944–958},
   url = {https://doi.org/10.1038/s41578-021-00336-6},
   year = {2021},
}

@article{Yazdani2023,
  title = {Hunting for {{Majoranas}}},
  author = {Yazdani, Ali and Von Oppen, Felix and Halperin, Bertrand I. and Yacoby, Amir},
  year = {2023},
  month = jun,
  journal = {Science},
  volume = {380},
  number = {6651},
  pages = {eade0850},
  issn = {0036-8075, 1095-9203},
  doi = {10.1126/science.ade0850},
  url = {https://www.science.org/doi/10.1126/science.ade0850}
}

@article{Rachel2025,
title = {Majorana quasiparticles in atomic spin chains on superconductors},
journal = {Phys. Rep.},
volume = {1099},
pages = {1-28},
year = {2025},
issn = {0370-1573},
doi = {https://doi.org/10.1016/j.physrep.2024.10.005},
url = {https://www.sciencedirect.com/science/article/pii/S0370157324003648},
author = {Stephan Rachel and Roland Wiesendanger},
}

@article{Mourik2012,
  title = {Signatures of {{Majorana Fermions}} in {{Hybrid Superconductor-Semiconductor Nanowire Devices}}},
  author = {Mourik, V. and Zuo, K. and Frolov, S. M. and Plissard, S. R. and Bakkers, E. P. A. M. and Kouwenhoven, L. P.},
  year = {2012},
  month = may,
  journal = {Science},
  volume = {336},
  number = {6084},
  pages = {1003--1007},
  issn = {0036-8075, 1095-9203},
  doi = {10.1126/science.1222360},
  url = {https://www.science.org/doi/10.1126/science.1222360}
}

@article{Das2012,
  title = {Zero-Bias Peaks and Splitting in an {{Al}}--{{InAs}} Nanowire Topological Superconductor as a Signature of {{Majorana}} Fermions},
  author = {Das, Anindya and Ronen, Yuval and Most, Yonatan and Oreg, Yuval and Heiblum, Moty and Shtrikman, Hadas},
  year = {2012},
  month = dec,
  journal = {Nat. Phys.},
  volume = {8},
  number = {12},
  pages = {887--895},
  issn = {1745-2473, 1745-2481},
  doi = {10.1038/nphys2479},
  url = {https://www.nature.com/articles/nphys2479}
}

@article{Deng2012,
  title = {Anomalous {{Zero-Bias Conductance Peak}} in a {{Nb}}--{{InSb Nanowire}}--{{Nb Hybrid Device}}},
  author = {Deng, M. T. and Yu, C. L. and Huang, G. Y. and Larsson, M. and Caroff, P. and Xu, H. Q.},
  year = {2012},
  month = dec,
  journal = {Nano Lett.},
  volume = {12},
  number = {12},
  pages = {6414--6419},
  issn = {1530-6984, 1530-6992},
  doi = {10.1021/nl303758w},
  url = {https://pubs.acs.org/doi/10.1021/nl303758w}
}

@article{Lee2012,
  title = {Zero-{{Bias Anomaly}} in a {{Nanowire Quantum Dot Coupled}} to {{Superconductors}}},
  author = {Lee, Eduardo J. H. and Jiang, Xiaocheng and Aguado, Ram{\'o}n and Katsaros, Georgios and Lieber, Charles M. and De Franceschi, Silvano},
  year = {2012},
  month = oct,
  journal = {Phys. Rev. Lett.},
  volume = {109},
  number = {18},
  pages = {186802},
  issn = {0031-9007, 1079-7114},
  doi = {10.1103/PhysRevLett.109.186802},
  url = {https://link.aps.org/doi/10.1103/PhysRevLett.109.186802}
}

@article{Churchill2013,
  title = {Superconductor-Nanowire Devices from Tunneling to the Multichannel Regime: {{Zero-bias}} Oscillations and Magnetoconductance Crossover},
  author = {Churchill, H. O. H. and Fatemi, V. and {Grove-Rasmussen}, K. and Deng, M. T. and Caroff, P. and Xu, H. Q. and Marcus, C. M.},
  year = {2013},
  month = jun,
  journal = {Phys. Rev. B},
  volume = {87},
  number = {24},
  pages = {241401},
  issn = {1098-0121, 1550-235X},
  doi = {10.1103/PhysRevB.87.241401},
  url = {https://link.aps.org/doi/10.1103/PhysRevB.87.241401}
}

@article{Finck2013,
  title = {Anomalous {{Modulation}} of a {{Zero-Bias Peak}} in a {{Hybrid Nanowire-Superconductor Device}}},
  author = {Finck, A. D. K. and Van Harlingen, D. J. and Mohseni, P. K. and Jung, K. and Li, X.},
  year = {2013},
  month = mar,
  journal = {Phys. Rev. Lett.},
  volume = {110},
  number = {12},
  pages = {126406},
  issn = {0031-9007, 1079-7114},
  doi = {10.1103/PhysRevLett.110.126406},
  url = {https://link.aps.org/doi/10.1103/PhysRevLett.110.126406}
}

@article{Nichele2017,
  title = {Scaling of {{Majorana Zero-Bias Conductance Peaks}}},
  author = {Nichele, Fabrizio and Drachmann, Asbj{\o}rn C. C. and Whiticar, Alexander M. and O'Farrell, Eoin C. T. and Suominen, Henri J. and Fornieri, Antonio and Wang, Tian and Gardner, Geoffrey C. and Thomas, Candice and Hatke, Anthony T. and Krogstrup, Peter and Manfra, Michael J. and Flensberg, Karsten and Marcus, Charles M.},
  year = {2017},
  month = sep,
  journal = {Phys. Rev. Lett.},
  volume = {119},
  number = {13},
  pages = {136803},
  issn = {0031-9007, 1079-7114},
  doi = {10.1103/PhysRevLett.119.136803},
  url = {https://link.aps.org/doi/10.1103/PhysRevLett.119.136803}
}

@article{Deng2018,
  title = {Nonlocality of {{Majorana}} Modes in Hybrid Nanowires},
  author = {Deng, M.-T. and Vaitiek{\.e}nas, S. and Prada, E. and {San-Jose}, P. and Nyg{\aa}rd, J. and Krogstrup, P. and Aguado, R. and Marcus, C. M.},
  year = {2018},
  month = aug,
  journal = {Phys. Rev. B},
  volume = {98},
  number = {8},
  pages = {085125},
  issn = {2469-9950, 2469-9969},
  doi = {10.1103/PhysRevB.98.085125},
  url = {https://link.aps.org/doi/10.1103/PhysRevB.98.085125}
}

@article{Gul2018,
  title = {Ballistic {{Majorana}} Nanowire Devices},
  author = {G{\"u}l, {\"O}nder and Zhang, Hao and Bommer, Jouri D. S. and De Moor, Michiel W. A. and Car, Diana and Plissard, S{\'e}bastien R. and Bakkers, Erik P. A. M. and Geresdi, Attila and Watanabe, Kenji and Taniguchi, Takashi and Kouwenhoven, Leo P.},
  year = {2018},
  month = mar,
  journal = {Nature Nanotech},
  volume = {13},
  number = {3},
  pages = {192--197},
  issn = {1748-3387, 1748-3395},
  doi = {10.1038/s41565-017-0032-8},
  url = {https://www.nature.com/articles/s41565-017-0032-8}
}

@misc{Zhang2021,
      title = {Large Zero-Bias Peaks in {{InSb-Al}} Hybrid Semiconductor-Superconductor Nanowire Devices},
      author = {Zhang, Hao and {de Moor}, Michiel W.A. and Bommer, Jouri D.S. and Xu, Di and Wang, Guanzhong and {van Loo}, Nick and Liu, Chun-Xiao and Gazibegovic, Sasa and Logan, John A. and Car, Diana and {Op het Veld}, Roy L. M. and {van Veldhoven}, Petrus J. and Koellinga, Sebastian and Verheijen, Marcel A. and Pendharkar, Mihir and Pennachio, Daniel J. and Shojaei, Borzoyeh and Lee, Joon Sue and Palmstr{\o}m, Chris J. and Bakkers, Erik P.A.M. and Das Sarma, S. and Kouwenhoven, Leo P.},
      year = {2021},
      eprint={2101.11456},
      archivePrefix={arXiv},
      primaryClass={cond-mat.mes-hall},
      url={https://arxiv.org/abs/2101.11456}
}

@article{Menard2020,
  title = {Conductance-{{Matrix Symmetries}} of a {{Three-Terminal Hybrid Device}}},
  author = {M{\'e}nard, G. C. and Anselmetti, G. L. R. and Martinez, E. A. and Puglia, D. and Malinowski, F. K. and Lee, J. S. and Choi, S. and Pendharkar, M. and Palmstr{\o}m, C. J. and Flensberg, K. and Marcus, C. M. and Casparis, L. and Higginbotham, A. P.},
  year = {2020},
  month = jan,
  journal = {Phys. Rev. Lett.},
  volume = {124},
  number = {3},
  pages = {036802},
  issn = {0031-9007, 1079-7114},
  doi = {10.1103/PhysRevLett.124.036802},
  url = {https://link.aps.org/doi/10.1103/PhysRevLett.124.036802}
}

@article{Heedt2021,
  title = {Shadow-Wall Lithography of Ballistic Superconductor--Semiconductor Quantum Devices},
  author = {Heedt, Sebastian and {Quintero-P{\'e}rez}, Marina and Borsoi, Francesco and Fursina, Alexandra and Van Loo, Nick and Mazur, Grzegorz P. and Nowak, Micha{\l} P. and Ammerlaan, Mark and Li, Kongyi and Korneychuk, Svetlana and Shen, Jie and Van De Poll, May An Y. and Badawy, Ghada and Gazibegovic, Sasa and De Jong, Nick and Aseev, Pavel and Van Hoogdalem, Kevin and Bakkers, Erik P. A. M. and Kouwenhoven, Leo P.},
  year = {2021},
  month = aug,
  journal = {Nat. Commun.},
  volume = {12},
  number = {1},
  pages = {4914},
  issn = {2041-1723},
  doi = {10.1038/s41467-021-25100-w},
  url = {https://www.nature.com/articles/s41467-021-25100-w}
}

@article{Puglia2021,
  title = {Closing of the Induced Gap in a Hybrid Superconductor-Semiconductor Nanowire},
  author = {Puglia, D. and Martinez, E. A. and M{\'e}nard, G. C. and P{\"o}schl, A. and Gronin, S. and Gardner, G. C. and Kallaher, R. and Manfra, M. J. and Marcus, C. M. and Higginbotham, A. P. and Casparis, L.},
  year = {2021},
  month = jun,
  journal = {Phys. Rev. B},
  volume = {103},
  number = {23},
  pages = {235201},
  issn = {2469-9950, 2469-9969},
  doi = {10.1103/PhysRevB.103.235201},
  url = {https://link.aps.org/doi/10.1103/PhysRevB.103.235201}
}

@article{Aghaee2023,
  title = {{{InAs-Al}} Hybrid Devices Passing the Topological Gap Protocol},
  author = {Aghaee, Morteza and others},
  year = {2023},
  month = jun,
  journal = {Phys. Rev. B},
  volume = {107},
  number = {24},
  pages = {245423},
  doi = {10.1103/PhysRevB.107.245423}
}

@article{Aghaee2025a,
  title = {Interferometric Single-Shot Parity Measurement in {{InAs}}--{{Al}} Hybrid Devices},
  author = {Aghaee, Morteza and others},
  year = {2025},
  month = feb,
  journal = {Nature},
  volume = {638},
  number = {8051},
  pages = {651--655},
  doi = {10.1038/s41586-024-08445-2},
  url = {https://www.nature.com/articles/s41586-024-08445-2}
}

@article{Rokhinson2012,
  title = {The Fractional a.c. {{Josephson}} Effect in a Semiconductor--Superconductor Nanowire as a Signature of {{Majorana}} Particles},
  author = {Rokhinson, Leonid P. and Liu, Xinyu and Furdyna, Jacek K.},
  year = {2012},
  month = nov,
  journal = {Nat. Phys.},
  volume = {8},
  number = {11},
  pages = {795--799},
  issn = {1745-2473, 1745-2481},
  doi = {10.1038/nphys2429},
  url = {https://www.nature.com/articles/nphys2429}
}

@article{Laroche2019,
  title = {Observation of the 4{$\pi$}-Periodic {{Josephson}} Effect in Indium Arsenide Nanowires},
  author = {Laroche, Dominique and Bouman, Dani{\"e}l and Van Woerkom, David J. and Proutski, Alex and Murthy, Chaitanya and Pikulin, Dmitry I. and Nayak, Chetan and Van Gulik, Ruben J. J. and Nyg{\aa}rd, Jesper and Krogstrup, Peter and Kouwenhoven, Leo P. and Geresdi, Attila},
  year = {2019},
  month = jan,
  journal = {Nat. Commun.},
  volume = {10},
  number = {1},
  pages = {245},
  issn = {2041-1723},
  doi = {10.1038/s41467-018-08161-2},
  url = {https://www.nature.com/articles/s41467-018-08161-2}
}

@article{Higginbotham2015,
  title = {Parity Lifetime of Bound States in a Proximitized Semiconductor Nanowire},
  author = {Higginbotham, A.~P. and Albrecht, S.~M. and Kir{\v s}anskas, G. and Chang, W. and Kuemmeth, F. and Krogstrup, P. and Jespersen, T.~S. and Nyg{\aa}rd, J. and Flensberg, K. and Marcus, C.~M.},
  year = {2015},
  month = dec,
  journal = {Nat. Phys.},
  volume = {11},
  number = {12},
  pages = {1017--1021},
  issn = {1745-2473, 1745-2481},
  doi = {10.1038/nphys3461},
  url = {https://www.nature.com/articles/nphys3461}
}

@article{Albrecht2016,
  title = {Exponential Protection of Zero Modes in {{Majorana}} Islands},
  author = {Albrecht, S. M. and Higginbotham, A. P. and Madsen, M. and Kuemmeth, F. and Jespersen, T. S. and Nyg{\aa}rd, J. and Krogstrup, P. and Marcus, C. M.},
  year = {2016},
  month = mar,
  journal = {Nature},
  volume = {531},
  number = {7593},
  pages = {206--209},
  issn = {0028-0836, 1476-4687},
  doi = {10.1038/nature17162},
  url = {https://www.nature.com/articles/nature17162}
}

@article{Deng2016,
  title = {Majorana Bound State in a Coupled Quantum-Dot Hybrid-Nanowire System},
  author = {Deng, M. T. and Vaitiek{\.e}nas, S. and Hansen, E. B. and Danon, J. and Leijnse, M. and Flensberg, K. and Nyg{\aa}rd, J. and Krogstrup, P. and Marcus, C. M.},
  year = {2016},
  month = dec,
  journal = {Science},
  volume = {354},
  number = {6319},
  pages = {1557--1562},
  issn = {0036-8075, 1095-9203},
  doi = {10.1126/science.aaf3961},
  url = {https://www.science.org/doi/10.1126/science.aaf3961}
}

@article{Albrecht2017,
  title = {Transport {{Signatures}} of {{Quasiparticle Poisoning}} in a {{Majorana Island}}},
  author = {Albrecht, S. M. and Hansen, E. B. and Higginbotham, A. P. and Kuemmeth, F. and Jespersen, T. S. and Nyg{\aa}rd, J. and Krogstrup, P. and Danon, J. and Flensberg, K. and Marcus, C. M.},
  year = {2017},
  month = mar,
  journal = {Phys. Rev. Lett.},
  volume = {118},
  number = {13},
  pages = {137701},
  issn = {0031-9007, 1079-7114},
  doi = {10.1103/PhysRevLett.118.137701},
  url = {https://link.aps.org/doi/10.1103/PhysRevLett.118.137701}
}

@article{Shen2018,
  title = {Parity Transitions in the Superconducting Ground State of Hybrid {{InSb}}--{{Al Coulomb}} Islands},
  author = {Shen, Jie and Heedt, Sebastian and Borsoi, Francesco and Van Heck, Bernard and Gazibegovic, Sasa and Op Het Veld, Roy L. M. and Car, Diana and Logan, John A. and Pendharkar, Mihir and Ramakers, Senja J. J. and Wang, Guanzhong and Xu, Di and Bouman, Dani{\"e}l and Geresdi, Attila and Palmstr{\o}m, Chris J. and Bakkers, Erik P. A. M. and Kouwenhoven, Leo P.},
  year = {2018},
  month = nov,
  journal = {Nat. Commun.},
  volume = {9},
  number = {1},
  pages = {4801},
  issn = {2041-1723},
  doi = {10.1038/s41467-018-07279-7},
  url = {https://www.nature.com/articles/s41467-018-07279-7}
}

@article{Vaitiekenas2020,
  title = {Flux-Induced Topological Superconductivity in Full-Shell Nanowires},
  author = {Vaitiek{\.e}nas, S. and Winkler, G. W. and Van Heck, B. and Karzig, T. and Deng, M.-T. and Flensberg, K. and Glazman, L. I. and Nayak, C. and Krogstrup, P. and Lutchyn, R. M. and Marcus, C. M.},
  year = {2020},
  month = mar,
  journal = {Science},
  volume = {367},
  number = {6485},
  pages = {eaav3392},
  issn = {0036-8075, 1095-9203},
  doi = {10.1126/science.aav3392},
  url = {https://www.science.org/doi/10.1126/science.aav3392}
}

@article{VanZanten2020,
  title = {Photon-Assisted Tunnelling of Zero Modes in a {{Majorana}} Wire},
  author = {Van Zanten, David M. T. and Sabonis, Deividas and Suter, Judith and V{\"a}yrynen, Jukka I. and Karzig, Torsten and Pikulin, Dmitry I. and O'Farrell, Eoin C. T. and Razmadze, Davydas and Petersson, Karl D. and Krogstrup, Peter and Marcus, Charles M.},
  year = {2020},
  month = jun,
  journal = {Nat. Phys.},
  volume = {16},
  number = {6},
  pages = {663--668},
  issn = {1745-2473, 1745-2481},
  doi = {10.1038/s41567-020-0858-0},
  url = {https://www.nature.com/articles/s41567-020-0858-0}
}

@article{Rainis2012,
  title = {Majorana qubit decoherence by quasiparticle poisoning},
  author = {Rainis, Diego and Loss, Daniel},
  journal = {Phys. Rev. B},
  volume = {85},
  issue = {17},
  pages = {174533},
  numpages = {10},
  year = {2012},
  month = {May},
  publisher = {American Physical Society},
  doi = {10.1103/PhysRevB.85.174533},
  url = {https://link.aps.org/doi/10.1103/PhysRevB.85.174533}
}

@article{Knapp2018,
  title = {Dephasing of Majorana-based qubits},
  author = {Knapp, Christina and Karzig, Torsten and Lutchyn, Roman M. and Nayak, Chetan},
  journal = {Phys. Rev. B},
  volume = {97},
  issue = {12},
  pages = {125404},
  numpages = {14},
  year = {2018},
  month = {Mar},
  publisher = {American Physical Society},
  doi = {10.1103/PhysRevB.97.125404},
  url = {https://link.aps.org/doi/10.1103/PhysRevB.97.125404}
}

@article{Karzig2021,
  title = {{Quasiparticle Poisoning of Majorana Qubits}},
  author = {Karzig, Torsten and Cole, William S. and Pikulin, Dmitry I.},
  journal = {Phys. Rev. Lett.},
  volume = {126},
  issue = {5},
  pages = {057702},
  numpages = {7},
  year = {2021},
  month = {Feb},
  publisher = {American Physical Society},
  doi = {10.1103/PhysRevLett.126.057702},
  url = {https://link.aps.org/doi/10.1103/PhysRevLett.126.057702}
}

@article{Alicea2011,
   title = {Non-{A}belian statistics and topological quantum information processing in 1{D} wire networks},
   author = {Alicea, J and Oreg, Yuval and Refael, Gil and von Oppen, Felix and P. A. Fisher, Matthew},
   doi = {10.1038/nphys1915},
   journal = {Nat. Phys.},
   pages = {412-417},
   publisher = {Nature Publishing Group},
   volume = {7},
   url = {https://doi.org/10.1038/nphys1915},
   year = {2011},
}

@article{Clarke2011,
  title = {Majorana fermion exchange in quasi-one-dimensional networks},
  author = {Clarke, David J. and Sau, Jay D. and Tewari, Sumanta},
  journal = {Phys. Rev. B},
  volume = {84},
  issue = {3},
  pages = {035120},
  numpages = {8},
  year = {2011},
  month = {Jul},
  publisher = {American Physical Society},
  doi = {10.1103/PhysRevB.84.035120},
  url = {https://link.aps.org/doi/10.1103/PhysRevB.84.035120}
}

@article{Halperin2012,
  title = {Adiabatic manipulations of {Majorana} fermions in a three-dimensional network of quantum wires},
  author = {Halperin, Bertrand I. and Oreg, Yuval and Stern, Ady and Refael, Gil and Alicea, Jason and von Oppen, Felix},
  journal = {Phys. Rev. B},
  volume = {85},
  issue = {14},
  pages = {144501},
  numpages = {16},
  year = {2012},
  month = {Apr},
  publisher = {American Physical Society},
  doi = {10.1103/PhysRevB.85.144501},
  url = {https://link.aps.org/doi/10.1103/PhysRevB.85.144501}
}

@article{Bauer2019,
  title = {Topologically protected braiding in a single wire using {Floquet Majorana} modes},
  author = {Bauer, Bela and Pereg-Barnea, T. and Karzig, Torsten and Rieder, Maria-Theresa and Refael, Gil and Berg, Erez and Oreg, Yuval},
  journal = {Phys. Rev. B},
  volume = {100},
  issue = {4},
  pages = {041102},
  numpages = {6},
  year = {2019},
  month = {Jul},
  publisher = {American Physical Society},
  doi = {10.1103/PhysRevB.100.041102},
  url = {https://link.aps.org/doi/10.1103/PhysRevB.100.041102}
}

@article{Flensberg2011,
  title = {{Non-Abelian Operations on Majorana Fermions via Single-Charge Control}},
  author = {Flensberg, Karsten},
  journal = {Phys. Rev. Lett.},
  volume = {106},
  issue = {9},
  pages = {090503},
  numpages = {4},
  year = {2011},
  month = {Mar},
  publisher = {American Physical Society},
  doi = {10.1103/PhysRevLett.106.090503},
  url = {https://link.aps.org/doi/10.1103/PhysRevLett.106.090503}
}

@article{Sau2011,
  title = {Controlling non-{Abelian} statistics of {Majorana} fermions in semiconductor nanowires},
  author = {Sau, Jay D. and Clarke, David J. and Tewari, Sumanta},
  journal = {Phys. Rev. B},
  volume = {84},
  issue = {9},
  pages = {094505},
  numpages = {8},
  year = {2011},
  month = {Sep},
  publisher = {American Physical Society},
  doi = {10.1103/PhysRevB.84.094505},
  url = {https://link.aps.org/doi/10.1103/PhysRevB.84.094505}
}

@article{vanHeck2012,
doi = {10.1088/1367-2630/14/3/035019},
url = {https://dx.doi.org/10.1088/1367-2630/14/3/035019},
year = {2012},
month = {mar},
publisher = {IOP Publishing},
volume = {14},
number = {3},
pages = {035019},
author = {van Heck, B and Akhmerov, A R and Hassler, F and Burrello, M and Beenakker, C W J},
title = {Coulomb-assisted braiding of {Majorana} fermions in a {Josephson} junction array},
journal = {New J. Phys.},
}

@article{Hyart2013,
  title = {Flux-controlled quantum computation with {Majorana} fermions},
  author = {Hyart, T. and van Heck, B. and Fulga, I. C. and Burrello, M. and Akhmerov, A. R. and Beenakker, C. W. J.},
  journal = {Phys. Rev. B},
  volume = {88},
  issue = {3},
  pages = {035121},
  numpages = {17},
  year = {2013},
  month = {Jul},
  publisher = {American Physical Society},
  doi = {10.1103/PhysRevB.88.035121},
  url = {https://link.aps.org/doi/10.1103/PhysRevB.88.035121}
}

@article{Aasen2016,
  title = {{Milestones Toward Majorana-Based Quantum Computing}},
  author = {Aasen, David and Hell, Michael and Mishmash, Ryan V. and Higginbotham, Andrew and Danon, Jeroen and Leijnse, Martin and Jespersen, Thomas S. and Folk, Joshua A. and Marcus, Charles M. and Flensberg, Karsten and Alicea, Jason},
  journal = {Phys. Rev. X},
  volume = {6},
  issue = {3},
  pages = {031016},
  numpages = {28},
  year = {2016},
  month = {Aug},
  publisher = {American Physical Society},
  doi = {10.1103/PhysRevX.6.031016},
  url = {https://link.aps.org/doi/10.1103/PhysRevX.6.031016}
}

@article{Malciu2018,
  title = {Braiding {Majorana} zero modes using quantum dots},
  author = {Malciu, Corneliu and Mazza, Leonardo and Mora, Christophe},
  journal = {Phys. Rev. B},
  volume = {98},
  issue = {16},
  pages = {165426},
  numpages = {14},
  year = {2018},
  month = {Oct},
  publisher = {American Physical Society},
  doi = {10.1103/PhysRevB.98.165426},
  url = {https://link.aps.org/doi/10.1103/PhysRevB.98.165426}
}

@article{Martin2020,
  title = {{Double Braiding Majoranas for Quantum Computing and Hamiltonian Engineering}},
  author = {Martin, Ivar and Agarwal, Kartiek},
  journal = {PRX Quantum},
  volume = {1},
  issue = {2},
  pages = {020324},
  numpages = {17},
  year = {2020},
  month = {Dec},
  publisher = {American Physical Society},
  doi = {10.1103/PRXQuantum.1.020324},
  url = {https://link.aps.org/doi/10.1103/PRXQuantum.1.020324}
}

@article{Bonderson2008,
  title = {{Measurement-Only Topological Quantum Computation}},
  author = {Bonderson, Parsa and Freedman, Michael and Nayak, Chetan},
  journal = {Phys. Rev. Lett.},
  volume = {101},
  issue = {1},
  pages = {010501},
  numpages = {4},
  year = {2008},
  month = {Jun},
  publisher = {American Physical Society},
  doi = {10.1103/PhysRevLett.101.010501},
  url = {https://link.aps.org/doi/10.1103/PhysRevLett.101.010501}
}

@article{Bonderson2013,
  title = {Measurement-only topological quantum computation via tunable interactions},
  author = {Bonderson, Parsa},
  journal = {Phys. Rev. B},
  volume = {87},
  issue = {3},
  pages = {035113},
  numpages = {9},
  year = {2013},
  month = {Jan},
  publisher = {American Physical Society},
  doi = {10.1103/PhysRevB.87.035113},
  url = {https://link.aps.org/doi/10.1103/PhysRevB.87.035113}
}

@article{Vijay2016,
  title = {Teleportation-based quantum information processing with {Majorana} zero modes},
  author = {Vijay, Sagar and Fu, Liang},
  journal = {Phys. Rev. B},
  volume = {94},
  issue = {23},
  pages = {235446},
  numpages = {9},
  year = {2016},
  month = {Dec},
  publisher = {American Physical Society},
  doi = {10.1103/PhysRevB.94.235446},
  url = {https://link.aps.org/doi/10.1103/PhysRevB.94.235446}
}

@article{Karzig2017,
  title = {Scalable designs for quasiparticle-poisoning-protected topological quantum computation with {Majorana} zero modes},
  author = {Karzig, Torsten and Knapp, Christina and Lutchyn, Roman M. and Bonderson, Parsa and Hastings, Matthew B. and Nayak, Chetan and Alicea, Jason and Flensberg, Karsten and Plugge, Stephan and Oreg, Yuval and Marcus, Charles M. and Freedman, Michael H.},
  journal = {Phys. Rev. B},
  volume = {95},
  issue = {23},
  pages = {235305},
  numpages = {32},
  year = {2017},
  month = {Jun},
  publisher = {American Physical Society},
  doi = {10.1103/PhysRevB.95.235305},
  url = {https://link.aps.org/doi/10.1103/PhysRevB.95.235305}
}

@article{Plugge2017,
title = {Majorana box qubits},
doi = {10.1088/1367-2630/aa54e1},
url = {https://dx.doi.org/10.1088/1367-2630/aa54e1},
year = {2017},
month = {jan},
publisher = {IOP Publishing},
volume = {19},
number = {1},
pages = {012001},
author = {Plugge, Stephan and Rasmussen, Asbjørn and Egger, Reinhold and Flensberg, Karsten},
journal = {New J. Phys.},
}

@article{Cheng2011,
  title = {Nonadiabatic effects in the braiding of non-{Abelian} anyons in topological superconductors},
  author = {Cheng, Meng and Galitski, Victor and Das Sarma, S.},
  journal = {Phys. Rev. B},
  volume = {84},
  issue = {10},
  pages = {104529},
  numpages = {11},
  year = {2011},
  month = {Sep},
  publisher = {American Physical Society},
  doi = {10.1103/PhysRevB.84.104529},
  url = {https://link.aps.org/doi/10.1103/PhysRevB.84.104529}
}

@article{Karzig2015,
  title = {Shortcuts to non-{Abelian} braiding},
  author = {Karzig, Torsten and Pientka, Falko and Refael, Gil and von Oppen, Felix},
  journal = {Phys. Rev. B},
  volume = {91},
  issue = {20},
  pages = {201102},
  numpages = {5},
  year = {2015},
  month = {May},
  publisher = {American Physical Society},
  doi = {10.1103/PhysRevB.91.201102},
  url = {https://link.aps.org/doi/10.1103/PhysRevB.91.201102}
}

@article{Amorim2015,
  title = {Majorana braiding dynamics in nanowires},
  author = {Amorim, C\'assio Sozinho and Ebihara, Kazuto and Yamakage, Ai and Tanaka, Yukio and Sato, Masatoshi},
  journal = {Phys. Rev. B},
  volume = {91},
  issue = {17},
  pages = {174305},
  numpages = {8},
  year = {2015},
  month = {May},
  publisher = {American Physical Society},
  doi = {10.1103/PhysRevB.91.174305},
  url = {https://link.aps.org/doi/10.1103/PhysRevB.91.174305}
}

@article{Knapp2016,
  title = {{The Nature and Correction of Diabatic Errors in Anyon Braiding}},
  author = {Knapp, Christina and Zaletel, Michael and Liu, Dong E. and Cheng, Meng and Bonderson, Parsa and Nayak, Chetan},
  journal = {Phys. Rev. X},
  volume = {6},
  issue = {4},
  pages = {041003},
  numpages = {38},
  year = {2016},
  month = {Oct},
  publisher = {American Physical Society},
  doi = {10.1103/PhysRevX.6.041003},
  url = {https://link.aps.org/doi/10.1103/PhysRevX.6.041003}
}

@article{Rahmani2017,
   author = {Armin Rahmani and Babak Seradjeh and Marcel Franz},
   doi = {10.1103/PhysRevB.96.075158},
   issn = {24699969},
   issue = {7},
   journal = {Phys. Rev. B},
   month = {8},
   pages = {075158},
   publisher = {American Physical Society},
   title = {Optimal diabatic dynamics of {M}ajorana-based quantum gates},
   volume = {96},
   url = {https://doi.org/10.1103/PhysRevB.96.075158},
   year = {2017},
}

@article{Sekania2017,
  title = {Braiding errors in interacting {Majorana} quantum wires},
  author = {Sekania, Michael and Plugge, Stephan and Greiter, Martin and Thomale, Ronny and Schmitteckert, Peter},
  journal = {Phys. Rev. B},
  volume = {96},
  issue = {9},
  pages = {094307},
  numpages = {10},
  year = {2017},
  month = {Sep},
  publisher = {American Physical Society},
  doi = {10.1103/PhysRevB.96.094307},
  url = {https://link.aps.org/doi/10.1103/PhysRevB.96.094307}
}

@article{Zhang2019,
  title = {Effects of decoherence on diabatic errors in {Majorana} braiding},
  author = {Zhang, Zhen-Tao and Mei, Feng and Meng, Xiang-Guo and Liang, Bao-Long and Yang, Zhen-Shan},
  journal = {Phys. Rev. A},
  volume = {100},
  issue = {1},
  pages = {012324},
  numpages = {7},
  year = {2019},
  month = {Jul},
  publisher = {American Physical Society},
  doi = {10.1103/PhysRevA.100.012324},
  url = {https://link.aps.org/doi/10.1103/PhysRevA.100.012324}
}

@article{Nag2019,
  title = {Diabatic errors in {Majorana} braiding with bosonic bath},
  author = {Nag, Amit and Sau, Jay D.},
  journal = {Phys. Rev. B},
  volume = {100},
  issue = {1},
  pages = {014511},
  numpages = {14},
  year = {2019},
  month = {Jul},
  publisher = {American Physical Society},
  doi = {10.1103/PhysRevB.100.014511},
  url = {https://link.aps.org/doi/10.1103/PhysRevB.100.014511}
}

@article{Harper2019,
  title = {Majorana braiding in realistic nanowire Y-junctions and tuning forks},
  author = {Harper, Fenner and Pushp, Aakash and Roy, Rahul},
  journal = {Phys. Rev. Res.},
  volume = {1},
  issue = {3},
  pages = {033207},
  numpages = {16},
  year = {2019},
  month = {Dec},
  publisher = {American Physical Society},
  doi = {10.1103/PhysRevResearch.1.033207},
  url = {https://link.aps.org/doi/10.1103/PhysRevResearch.1.033207}
}

@article{Sanno2021,
  title = {Ab initio simulation of non-{Abelian} braiding statistics in topological superconductors},
  author = {Sanno, Takumi and Miyazaki, Shunsuke and Mizushima, Takeshi and Fujimoto, Satoshi},
  journal = {Phys. Rev. B},
  volume = {103},
  issue = {5},
  pages = {054504},
  numpages = {17},
  year = {2021},
  month = {Feb},
  publisher = {American Physical Society},
  doi = {10.1103/PhysRevB.103.054504},
  url = {https://link.aps.org/doi/10.1103/PhysRevB.103.054504}
}

@article{Xu2023,
  title = {Dynamics simulation of braiding two {Majorana} zero modes via a quantum dot},
  author = {Xu, Luting and Bai, Jing and Feng, Wei and Li, Xin-Qi},
  journal = {Phys. Rev. B},
  volume = {108},
  issue = {11},
  pages = {115411},
  numpages = {9},
  year = {2023},
  month = {Sep},
  publisher = {American Physical Society},
  doi = {10.1103/PhysRevB.108.115411},
  url = {https://link.aps.org/doi/10.1103/PhysRevB.108.115411}
}

@article{Mascot2023,
  title = {{Many-Body Majorana Braiding without an Exponential Hilbert Space}},
  author = {Mascot, Eric and Hodge, Themba and Crawford, Dan and Bedow, Jasmin and Morr, Dirk K. and Rachel, Stephan},
  journal = {Phys. Rev. Lett.},
  volume = {131},
  issue = {17},
  pages = {176601},
  numpages = {6},
  year = {2023},
  month = {Oct},
  publisher = {American Physical Society},
  doi = {10.1103/PhysRevLett.131.176601},
  url = {https://link.aps.org/doi/10.1103/PhysRevLett.131.176601}
}

@misc{Maciazek2023,
      title={Optimising the exchange of {Majorana} zero modes in a quantum nanowire network}, 
      author={Tomasz Maciazek and Aaron Conlon},
      year={2023},
      eprint={2310.13634},
      archivePrefix={arXiv},
      primaryClass={cond-mat.mes-hall},
      url={https://arxiv.org/abs/2310.13634}, 
}

@article{Boross2024,
  title = {Braiding-based quantum control of a {Majorana} qubit built from quantum dots},
  author = {Boross, P\'eter and P\'alyi, Andr\'as},
  journal = {Phys. Rev. B},
  volume = {109},
  issue = {12},
  pages = {125410},
  numpages = {14},
  year = {2024},
  month = {Mar},
  publisher = {American Physical Society},
  doi = {10.1103/PhysRevB.109.125410},
  url = {https://link.aps.org/doi/10.1103/PhysRevB.109.125410}
}

@article{Peeters2024,
  title = {Effect of impurities and disorder on the braiding dynamics of {Majorana} zero modes},
  author = {Peeters, Cole and Hodge, Themba and Mascot, Eric and Rachel, Stephan},
  journal = {Phys. Rev. B},
  volume = {110},
  issue = {21},
  pages = {214506},
  numpages = {9},
  year = {2024},
  month = {Dec},
  publisher = {American Physical Society},
  doi = {10.1103/PhysRevB.110.214506},
  url = {https://link.aps.org/doi/10.1103/PhysRevB.110.214506}
}

@article{Hodge2025,
  title = {{Characterizing Dynamic Hybridization of Majorana Zero Modes for Universal Quantum Computing}},
  author = {Hodge, Themba and Mascot, Eric and Crawford, Dan and Rachel, Stephan},
  journal = {Phys. Rev. Lett.},
  volume = {134},
  issue = {9},
  pages = {096601},
  numpages = {6},
  year = {2025},
  month = {Mar},
  publisher = {American Physical Society},
  doi = {10.1103/PhysRevLett.134.096601},
  url = {https://link.aps.org/doi/10.1103/PhysRevLett.134.096601}
}

@article{Scheurer2013,
	title = {Nonadiabatic processes in {Majorana} qubit systems},
	volume = {88},
	copyright = {http://link.aps.org/licenses/aps-default-license},
	issn = {1098-0121, 1550-235X},
	url = {https://link.aps.org/doi/10.1103/PhysRevB.88.064515},
	doi = {10.1103/PhysRevB.88.064515},
	number = {6},
	urldate = {2024-09-04},
	journal = {Phys. Rev. B},
	author = {Scheurer, M. S. and Shnirman, A.},
	month = aug,
	year = {2013},
	pages = {064515},
}

@article{Karzig2015_2,
	title = {Optimal control of {Majorana} zero modes},
	volume = {91},
	copyright = {http://link.aps.org/licenses/aps-default-license},
	issn = {1098-0121, 1550-235X},
	url = {https://link.aps.org/doi/10.1103/PhysRevB.91.201404},
	doi = {10.1103/PhysRevB.91.201404},
	number = {20},
	urldate = {2024-09-04},
	journal = {Phys. Rev. B},
	author = {Karzig, Torsten and Rahmani, Armin and Von Oppen, Felix and Refael, Gil},
	month = may,
	year = {2015},
	pages = {201404},
}

@article{Bauer2018,
	title = {Dynamics of {Majorana}-based qubits operated with an array of tunable gates},
	volume = {5},
	issn = {2542-4653},
	url = {https://scipost.org/10.21468/SciPostPhys.5.1.004},
	doi = {10.21468/SciPostPhys.5.1.004},
	number = {1},
	urldate = {2024-09-04},
	journal = {SciPost Phys.},
	author = {Bauer, Bela and Karzig, Torsten and Mishmash, Ryan and Antipov, Andrey and Alicea, Jason},
	month = jul,
	year = {2018},
	pages = {004},
}

@article{Conlon2019,
	title = {Error generation and propagation in {Majorana}-based topological qubits},
	volume = {100},
	issn = {2469-9950, 2469-9969},
	url = {https://link.aps.org/doi/10.1103/PhysRevB.100.134307},
	doi = {10.1103/PhysRevB.100.134307},
	number = {13},
	urldate = {2024-09-04},
	journal = {Phys. Rev. B},
	author = {Conlon, A. and Pellegrino, D. and Slingerland, J. K. and Dooley, S. and Kells, G.},
	month = oct,
	year = {2019},
	pages = {134307},
}

@article{Coopmans2021,
	title = {Protocol {Discovery} for the {Quantum} {Control} of {Majoranas} by {Differentiable} {Programming} and {Natural} {Evolution} {Strategies}},
	volume = {2},
	issn = {2691-3399},
	url = {https://link.aps.org/doi/10.1103/PRXQuantum.2.020332},
	doi = {10.1103/PRXQuantum.2.020332},
	number = {2},
	urldate = {2024-09-04},
	journal = {PRX Quantum},
	author = {Coopmans, Luuk and Luo, Di and Kells, Graham and Clark, Bryan K. and Carrasquilla, Juan},
	month = jun,
	year = {2021},
	pages = {020332},
}

@article{Xu2022,
	title = {Transport probe of the nonadiabatic transition caused by moving {Majorana} zero modes},
	volume = {105},
	issn = {2469-9950, 2469-9969},
	url = {https://link.aps.org/doi/10.1103/PhysRevB.105.245410},
	doi = {10.1103/PhysRevB.105.245410},
	number = {24},
	urldate = {2024-09-04},
	journal = {Phys. Rev. B},
	author = {Xu, Luting and Li, Xin-Qi},
	month = jun,
	year = {2022},
	pages = {245410},
}

@article{Truong2023,
	title = {Optimizing the transport of {Majorana} zero modes in one-dimensional topological superconductors},
	volume = {107},
	issn = {2469-9950, 2469-9969},
	url = {https://link.aps.org/doi/10.1103/PhysRevB.107.104516},
	doi = {10.1103/PhysRevB.107.104516},
	number = {10},
	urldate = {2024-09-04},
	journal = {Phys. Rev. B},
	author = {Truong, Bill P. and Agarwal, Kartiek and Pereg-Barnea, T.},
	month = mar,
	year = {2023},
	pages = {104516},
}

@article{Wang2024,
	title = {Transport and fusion of {Majorana} zero modes in the presence of nonadiabatic transitions},
	volume = {110},
	issn = {2469-9950, 2469-9969},
	url = {https://link.aps.org/doi/10.1103/PhysRevB.110.115402},
	doi = {10.1103/PhysRevB.110.115402},
	number = {11},
	urldate = {2025-03-09},
	journal = {Phys. Rev. B},
	author = {Wang, Qiongyao and Bai, Jing and Xu, Luting and Feng, Wei and Li, Xin-Qi},
	month = sep,
	year = {2024},
	pages = {115402},
}

@article{Sahu2024,
  title = {Transport of Majorana bound states in the presence of telegraph noise},
  author = {Sahu, Dibyajyoti and Gangadharaiah, Suhas},
  journal = {Phys. Rev. B},
  volume = {111},
  issue = {23},
  pages = {235306},
  numpages = {13},
  year = {2025},
  month = {Jun},
  publisher = {American Physical Society},
  doi = {10.1103/lr2b-nmrk},
  url = {https://link.aps.org/doi/10.1103/lr2b-nmrk}
}

@article{Pandey2025,
  title = {Diabatic error and propagation of {Majorana} zero modes in interacting quantum dots systems},
  author = {Pandey, Bradraj and Gupta, Gaurav Kumar and Alvarez, Gonzalo and Okamoto, Satoshi and Dagotto, Elbio},
  journal = {Phys. Rev. B},
  volume = {111},
  issue = {10},
  pages = {104311},
  numpages = {10},
  year = {2025},
  month = {Mar},
  publisher = {American Physical Society},
  doi = {10.1103/PhysRevB.111.104311},
  url = {https://link.aps.org/doi/10.1103/PhysRevB.111.104311}
}

@article{Brouwer2011,
  title = {{Probability Distribution of Majorana End-State Energies in Disordered Wires}},
  author = {Brouwer, Piet W. and Duckheim, Mathias and Romito, Alessandro and von Oppen, Felix},
  journal = {Phys. Rev. Lett.},
  volume = {107},
  issue = {19},
  pages = {196804},
  numpages = {4},
  year = {2011},
  month = {Nov},
  publisher = {American Physical Society},
  doi = {10.1103/PhysRevLett.107.196804},
  url = {https://link.aps.org/doi/10.1103/PhysRevLett.107.196804}
}

@article{Cai2013,
  title = {{Topological Superconductor to Anderson Localization Transition in One-Dimensional Incommensurate Lattices}},
  author = {Cai, Xiaoming and Lang, Li-Jun and Chen, Shu and Wang, Yupeng},
  journal = {Phys. Rev. Lett.},
  volume = {110},
  issue = {17},
  pages = {176403},
  numpages = {5},
  year = {2013},
  month = {Apr},
  publisher = {American Physical Society},
  doi = {10.1103/PhysRevLett.110.176403},
  url = {https://link.aps.org/doi/10.1103/PhysRevLett.110.176403}
}

@article{Hegde2016,
  title = {Majorana wave-function oscillations, fermion parity switches, and disorder in {Kitaev} chains},
  author = {Hegde, Suraj S. and Vishveshwara, Smitha},
  journal = {Phys. Rev. B},
  volume = {94},
  issue = {11},
  pages = {115166},
  numpages = {19},
  year = {2016},
  month = {Sep},
  publisher = {American Physical Society},
  doi = {10.1103/PhysRevB.94.115166},
  url = {https://link.aps.org/doi/10.1103/PhysRevB.94.115166}
}

@article{Boross2022,
  title = {Dephasing of {Majorana} qubits due to quasistatic disorder},
  author = {Boross, P\'eter and P\'alyi, Andr\'as},
  journal = {Phys. Rev. B},
  volume = {105},
  issue = {3},
  pages = {035413},
  numpages = {17},
  year = {2022},
  month = {Jan},
  publisher = {American Physical Society},
  doi = {10.1103/PhysRevB.105.035413},
  url = {https://link.aps.org/doi/10.1103/PhysRevB.105.035413}
}

@article{Paladino2014,
  title = {$1/f$ noise: Implications for solid-state quantum information},
  author = {Paladino, E. and Galperin, Y. M. and Falci, G. and Altshuler, B. L.},
  journal = {Rev. Mod. Phys.},
  volume = {86},
  issue = {2},
  pages = {361--418},
  numpages = {58},
  year = {2014},
  month = {Apr},
  publisher = {American Physical Society},
  doi = {10.1103/RevModPhys.86.361},
  url = {https://link.aps.org/doi/10.1103/RevModPhys.86.361}
}

@article{Malla2017,
  title = {Suppression of the Landau-Zener transition probability by weak classical noise},
  author = {Malla, Rajesh K. and Mishchenko, E. G. and Raikh, M. E.},
  journal = {Phys. Rev. B},
  volume = {96},
  issue = {7},
  pages = {075419},
  numpages = {8},
  year = {2017},
  month = {Aug},
  publisher = {American Physical Society},
  doi = {10.1103/PhysRevB.96.075419},
  url = {https://link.aps.org/doi/10.1103/PhysRevB.96.075419}
}

@article{Krzywda2020,
  title = {Adiabatic electron charge transfer between two quantum dots in presence of $1/f$ noise},
  author = {Krzywda, Jan A. and Cywi\ifmmode \acute{n}\else \'{n}\fi{}ski, \L{}ukasz},
  journal = {Phys. Rev. B},
  volume = {101},
  issue = {3},
  pages = {035303},
  numpages = {12},
  year = {2020},
  month = {Jan},
  publisher = {American Physical Society},
  doi = {10.1103/PhysRevB.101.035303},
  url = {https://link.aps.org/doi/10.1103/PhysRevB.101.035303}
}

@article{Mishmash2020,
  title = {Dephasing and leakage dynamics of noisy Majorana-based qubits: Topological versus Andreev},
  author = {Mishmash, Ryan V. and Bauer, Bela and von Oppen, Felix and Alicea, Jason},
  journal = {Phys. Rev. B},
  volume = {101},
  issue = {7},
  pages = {075404},
  numpages = {21},
  year = {2020},
  month = {Feb},
  publisher = {American Physical Society},
  doi = {10.1103/PhysRevB.101.075404},
  url = {https://link.aps.org/doi/10.1103/PhysRevB.101.075404}
}

@article{Wimmer2012,
   author = {M. Wimmer},
   issn = {00983500},
   issue = {4},
   journal = {ACM Trans. Math. Softw.},
   month = {8},
   title = {Algorithm 923: {E}fficient {N}umerical {C}omputation of the {P}faffian for {D}ense and {B}anded {S}kew-{S}ymmetric {M}atrices},
   volume = {38},
   url = {https://doi.org/10.1145/2331130.2331138},
   year = {2012},
}

@article{Bravyi2017,
   author = {Bravyi, S and Gosset, D},
   doi = {10.1007/s00220-017-2976-9},
   issn = {14320916},
   issue = {2},
   journal = {Commun. Math. Phys.},
   month = {12},
   pages = {451-500},
   publisher = {Springer New York LLC},
   title = {Complexity of Quantum Impurity Problems},
   volume = {356},
   url = {https://doi.org/10.1007/s00220-017-2976-9},
   year = {2017},
}

@article{DeMoor2018,
  title = {Electric Field Tunable Superconductor-Semiconductor Coupling in {{Majorana}} Nanowires},
  author = {De Moor, Michiel W A and Bommer, Jouri D S and Xu, Di and Winkler, Georg W and Antipov, Andrey E and Bargerbos, Arno and Wang, Guanzhong and Loo, Nick Van and Op Het Veld, Roy L M and Gazibegovic, Sasa and Car, Diana and Logan, John A and Pendharkar, Mihir and Lee, Joon Sue and M Bakkers, Erik P A and Palmstr{\o}m, Chris J and Lutchyn, Roman M and Kouwenhoven, Leo P and Zhang, Hao},
  year = {2018},
  month = oct,
  journal = {New J. Phys.},
  volume = {20},
  number = {10},
  pages = {103049},
  issn = {1367-2630},
  doi = {10.1088/1367-2630/aae61d},
  url = {https://iopscience.iop.org/article/10.1088/1367-2630/aae61d}
}

@article{Vaitiekenas2018,
  title = {Effective g {{Factor}} of {{Subgap States}} in {{Hybrid Nanowires}}},
  author = {Vaitiek{\.e}nas, S. and Deng, M.-T. and Nyg{\aa}rd, J. and Krogstrup, P. and Marcus, C. M.},
  year = {2018},
  month = jul,
  journal = {Phys. Rev. Lett.},
  volume = {121},
  number = {3},
  pages = {037703},
  issn = {0031-9007, 1079-7114},
  doi = {10.1103/PhysRevLett.121.037703},
  url = {https://link.aps.org/doi/10.1103/PhysRevLett.121.037703}
}

@article{Bordin2024,
  title = {Crossed {{Andreev Reflection}} and {{Elastic Cotunneling}} in {{Three Quantum Dots Coupled}} by {{Superconductors}}},
  author = {Bordin, Alberto and Li, Xiang and Van Driel, David and Wolff, Jan Cornelis and Wang, Qingzhen and Ten Haaf, Sebastiaan L. D. and Wang, Guanzhong and Van Loo, Nick and Kouwenhoven, Leo P. and Dvir, Tom},
  year = 2024,
  month = feb,
  journal = {Phys. Rev. Lett.},
  volume = {132},
  number = {5},
  pages = {056602},
  issn = {0031-9007, 1079-7114},
  doi = {10.1103/PhysRevLett.132.056602}
}

@article{Bordin2025,
  title = {Enhanced {{Majorana}} Stability in a Three-Site {{Kitaev}} Chain},
  author = {Bordin, Alberto and Liu, Chun-Xiao and Dvir, Tom and Zatelli, Francesco and Ten Haaf, Sebastiaan L. D. and Van Driel, David and Wang, Guanzhong and Van Loo, Nick and Zhang, Yining and Wolff, Jan Cornelis and Van Caekenberghe, Thomas and Badawy, Ghada and Gazibegovic, Sasa and Bakkers, Erik P. A. M. and Wimmer, Michael and Kouwenhoven, Leo P. and Mazur, Grzegorz P.},
  year = 2025,
  month = jun,
  journal = {Nat. Nanotechnol.},
  volume = {20},
  number = {6},
  pages = {726--731},
  issn = {1748-3387, 1748-3395},
  doi = {10.1038/s41565-025-01894-4}
}

@article{Dvir2023,
  title = {Realization of a Minimal {{Kitaev}} Chain in Coupled Quantum Dots},
  author = {Dvir, Tom and Wang, Guanzhong and Van Loo, Nick and Liu, Chun-Xiao and Mazur, Grzegorz P. and Bordin, Alberto and Ten Haaf, Sebastiaan L. D. and Wang, Ji-Yin and Van Driel, David and Zatelli, Francesco and Li, Xiang and Malinowski, Filip K. and Gazibegovic, Sasa and Badawy, Ghada and Bakkers, Erik P. A. M. and Wimmer, Michael and Kouwenhoven, Leo P.},
  year = 2023,
  month = feb,
  journal = {Nature},
  volume = {614},
  number = {7948},
  pages = {445--450},
  issn = {0028-0836, 1476-4687},
  doi = {10.1038/s41586-022-05585-1}
}

@article{Mazur2022,
  title = {Spin-{{Mixing Enhanced Proximity Effect}} in {{Aluminum}}-{{Based Superconductor}}--{{Semiconductor Hybrids}}},
  author = {Mazur, Grzegorz P. and Van Loo, Nick and Wang, Ji-Yin and Dvir, Tom and Wang, Guanzhong and Khindanov, Aleksei and Korneychuk, Svetlana and Borsoi, Francesco and Dekker, Robin C. and Badawy, Ghada and Vinke, Peter and Gazibegovic, Sasa and Bakkers, Erik P. A. M. and P{\'e}rez, Marina Quintero- and Heedt, Sebastian and Kouwenhoven, Leo P.},
  year = 2022,
  month = aug,
  journal = {Adv. Mater.},
  volume = {34},
  number = {33},
  pages = {2202034},
  issn = {0935-9648, 1521-4095},
  doi = {10.1002/adma.202202034}
}

@article{Mazur2024,
  title = {Gate-Tunable {{Josephson}} Diode},
  author = {Mazur, G.P. and Van Loo, N. and Van Driel, D. and Wang, J.-Y. and Badawy, G. and Gazibegovic, S. and Bakkers, E.P.A.M. and Kouwenhoven, L.P.},
  year = 2024,
  month = nov,
  journal = {Phys. Rev. Applied},
  volume = {22},
  number = {5},
  pages = {054034},
  issn = {2331-7019},
  doi = {10.1103/PhysRevApplied.22.054034}
}

@article{Wang2022,
  title = {Singlet and Triplet {{Cooper}} Pair Splitting in Hybrid Superconducting Nanowires},
  author = {Wang, Guanzhong and Dvir, Tom and Mazur, Grzegorz P. and Liu, Chun-Xiao and Van Loo, Nick and Ten Haaf, Sebastiaan L. D. and Bordin, Alberto and Gazibegovic, Sasa and Badawy, Ghada and Bakkers, Erik P. A. M. and Wimmer, Michael and Kouwenhoven, Leo P.},
  year = 2022,
  month = dec,
  journal = {Nature},
  volume = {612},
  number = {7940},
  pages = {448--453},
  issn = {0028-0836, 1476-4687},
  doi = {10.1038/s41586-022-05352-2}
}

@article{Zatelli2024,
  title = {Robust Poor Man's {{Majorana}} Zero Modes Using {{Yu-Shiba-Rusinov}} States},
  author = {Zatelli, Francesco and Van Driel, David and Xu, Di and Wang, Guanzhong and Liu, Chun-Xiao and Bordin, Alberto and Roovers, Bart and Mazur, Grzegorz P. and Van Loo, Nick and Wolff, Jan C. and Bozkurt, A. Mert and Badawy, Ghada and Gazibegovic, Sasa and Bakkers, Erik P. A. M. and Wimmer, Michael and Kouwenhoven, Leo P. and Dvir, Tom},
  year = 2024,
  month = sep,
  journal = {Nat. Commun.},
  volume = {15},
  number = {1},
  pages = {7933},
  issn = {2041-1723},
  doi = {10.1038/s41467-024-52066-2}
}

@article{Garrido1962,
   author = {L. M. Garrido and F. J. Sancho},
   doi = {10.1016/0031-8914(62)90109-X},
   journal = {Physica},
   pages = {553-560},
   title = {Degree of {A}pproximate {V}alidity of the {A}diabatic {I}nvariance in {Q}uantum {M}echanics},
   volume = {28},
   url = {https://doi.org/10.1016/0031-8914(62)90109-X},
   year = {1962},
}

\end{document}